\documentclass[preprint,authoryear,12pt]{elsarticle}
\usepackage[top=0.5in, bottom=0.5in, left=0.5in, right=0.5in]{geometry}

\usepackage{epsfig}
\usepackage{marvosym}
\usepackage{subfigure}

\usepackage{amssymb}
\usepackage{booktabs}
\usepackage{multirow}
\usepackage[a4paper=true,%
breaklinks=true,%
colorlinks=true,%
pdfauthor={X. P. Deng et al.},%
pdftitle={Interplanetary spacecraft navigation using pulsars}%
]{hyperref}

 \usepackage{amssymb}
 \usepackage{amsthm}
 \usepackage{amsmath}
 \usepackage{longtable}
 \usepackage{lscape}
 \usepackage{hyperref}

 \usepackage{graphicx}
 \usepackage{natbib}
 \usepackage{float}
\journal{Advances in Space Research}

\begin{document}

\begin{frontmatter}



\title{Interplanetary spacecraft navigation using pulsars}


\author{X. P. Deng\corref{cor}\fnref{footnote2}}
\address{National Space Science Center, Chinese Academy of Sciences, Beijing~100190, China\\
CSIRO Astronomy and Space Science, Australia Telescope National Facility, P.O.~Box~76, Epping NSW~1710, Australia\\
School of Information Science and Engineering, Graduate University of Chinese Academy of Sciences, Beijing~100190, China}
\cortext[cor]{Corresponding author}
\ead{xinping.deng@gmail.com}


\author{G. Hobbs\fnref{footnote3}}
\address{CSIRO Astronomy and Space Science, Australia Telescope National Facility, P.O.~Box~76, Epping NSW~1710, Australia}
\ead{george.hobbs@csiro.au}

\author{X. P. You}
\address{School of Physical Science and Technology, Southwest University, Chongqing 400715, China}
\ead{yxp0910@swu.edu.cn}

\author{M. T. Li}
\address{National Space Science Center, Chinese Academy of Sciences, Beijing~100190, China}
\ead{limingtao@nssc.ac.cn}

\author{M. J. Keith}
\address{CSIRO Astronomy and Space Science, Australia Telescope National Facility, P.O.~Box~76, Epping NSW~1710, Australia}
\ead{mkeith@pulsarastronomy.net }

\author{R. M. Shannon}
\address{CSIRO Astronomy and Space Science, Australia Telescope National Facility, P.O.~Box~76, Epping NSW~1710, Australia}
\ead{ryan.shannon@csiro.au }

\author{W. Coles }
\address{Electrical and Computer Engineering, University of California at San Diego, La Jolla, California, U.S.A.}
\ead{bcoles@ucsd.edu}

\author{R. N. Manchester }
\address{CSIRO Astronomy and Space Science, Australia Telescope National Facility, P.O.~Box~76, Epping NSW~1710, Australia}
\ead{dick.manchester@csiro.au}

\author{J. H. Zheng}
\address{National Space Science Center, Chinese Academy of Sciences, Beijing~100190, China}
\ead{zhengjianhua@nssc.ac.cn}

\author{X. Z. Yu}
\address{National Space Science Center, Chinese Academy of Sciences, Beijing~100190, China}
\ead{yuxizheng@nssc.ac.cn}

\author{D. Gao}
\address{National Space Science Center, Chinese Academy of Sciences, Beijing~100190, China}
\ead{gaodong@nssc.ac.cn}

\author{X. Wu}
\address{National Space Science Center, Chinese Academy of Sciences, Beijing~100190, China}
\ead{wuxia@nssc.ac.cn}

\author{D. Chen}
\address{National Space Science Center, Chinese Academy of Sciences, Beijing~100190, China}
\ead{ding@nssc.ac.cn}

\begin{abstract}
We demonstrate how observations of pulsars can be used to help navigate a spacecraft travelling in the solar system. We make use of archival observations of millisecond pulsars from the Parkes radio telescope in order to demonstrate the effectiveness of the method and highlight issues, such as pulsar spin irregularities, which need to be accounted for.  We show that observations of four millisecond pulsars every seven days using a realistic X-ray telescope on the spacecraft throughout a journey from Earth to Mars can lead to position determinations better than $\sim$20\,km and velocity measurements with a precision of $\sim$0.1\,ms$^{-1}$.  
\end{abstract}


\end{frontmatter}


\section{Introduction}

Accurate navigation through the solar system is essential for existing and future spacecraft missions.  In this paper we describe algorithms that can be used to develop an autonomous navigation system based on X-ray observations of pulsars in our Galaxy.  

Existing methods for navigating spacecraft depend on the duration of the mission and the required altitude above Earth. Low-earth orbit satellites can be navigated with great precision using the Global Positioning Satellite (GPS) system \citep{axelrad1996gps}. \citet{yu2011real} analysed GPS data collected from the CHAllenging Minisatellite Payload  (CHAMP) satellite that was launched in the year 2000 for atmospheric and ionospheric research. It was shown that the off-line absolute position and velocity determination error is approximately 10\,m and 0.01\,mms$^{-1}$ respectively.  
However, navigating a spacecraft with the existing GPS system restricts the maximum altitude of the satellite to $\sim$\,20000\,km \citep{bauer1998spaceborne}.  Clearly this system cannot be used for interplanetary trajectories. 

It is possible to use an inertial navigation system (INS) to estimate a spacecraft's position, orientation and velocity \citep{lawrence1998outline}.  Such a system is based on  accelerometers and gyroscopes and does not make use of any information external to the spacecraft. The precision of the position determination diminishes with time and, for long missions, is commonly combined with other methods. 

For interplanetary navigation it is possible to use an on-board camera to obtain images of the sky. Such images can then be used to determine the position, velocity and orientation of the spacecraft.  \citet{bhaskaran2000deep} showed that such a system can be used for position determination with an accuracy of $\sim$\,250\,km and velocity determination to $\sim$\,0.2\,ms$^{-1}$.   

As part of the NRL-801 experiment for the Advanced Research and Global Observation Satellite (ARGOS), \citet{wood1993navigation} and  \citet{hanson2006principles} presented  a comprehensive study using X-ray sources to determine a spacecraft's attitude, position and time. The position of a spacecraft could be determined using the occultation of an X-ray source behind the Earth's or the Moon's limb. An accuracy of the order of tens of meters  and time determination to $\sim$\,30\,${\mu}$s over many years was predicted. 

Currently the most common means to track spacecraft on interplanetary orbits is to use a network of large radio telescopes.  This method does not provide an autonomous means to determine the spacecraft's position, the signal delay between the ground station and spacecraft increases with distance and the precision of the position determination decreases with distance. According to \citet{curkendall2013delta}, an inherent angular precision on the order of 10 nanoradians is achievable using the Deep Space Network.  This corresponds to a position accuracy of $\sim$1.5\,km for a distance of one astronomical unit from Earth (the positional accuracy scales linearly with distance).

\subsection{Using pulsars to navigate spacecraft}

The idea of using pulsar timing observations to navigate spacecraft was first presented by \citet{downs1974interplanetary}. Pulsars are rapidly rotating neutron stars which emit beams of electromagnetic radiation and, for those bright enough to be of interest for spacecraft navigation, lie within our Galaxy.  For a fortuitous line of sight,  at least one of the radiation beams will sweep across the Earth allowing the pulsar to be detectable by searching for periodic pulses of emission.  Pulsars are usually discovered and observed using large ground-based radio telescopes.  The exceptional pulse stability has led to numerous applications such as testing the general theory of relativity \citep{kramer2006tests}, developing a pulsar-based time standard \citep{hobbs2012development}, measuring the mass of solar system planets \citep{champion2010measuring} and searching for evidence of gravitational waves \citep{van2011placing}.  Such applications begin from measurements of pulse times-of-arrival (ToAs) at a specified observatory.  Software, such as the \textsc{tempo2} package \citep{G-Hobbs_2006, R-Edwards_2006}, is used to convert these ToAs to pulse arrival times at the solar system barycentre (SSB), hereafter known as ``barycentric arrival times'', using a solar-system ephemeris (e.g., the Jet Propulsion Laboratory's DE421 planetary ephemeris; \citealt{folkner2008planetary}). A model of the astrometric, pulse and orbital parameters for the pulsar, and the group delay in the plasma between the pulsar and Earth, is subsequently used to predict the barycentric arrival times. The differences between the actual and predicted barycentric arrival times are called ``timing residuals''. Timing residuals result from various causes: (1) errors in the ToA measurement; (2) errors in the pulsar timing model; (3) errors in the solar system ephemeris; (4) errors in the model of the interstellar plasma; (5) phenomena not included in the timing model such as the effects of gravitational waves or random fluctuations in the pulse emission time intrinsic to the pulsar. In traditional pulsar timing analysis a linear\footnote{The timing model is not fully linear. Some of the parameters, such as the pulsar's pulse period, are linear.  Other parameters are linearised in \textsc{tempo2} and rely on having a reasonable estimate of the parameter.} least-squares-fitting procedure is carried out to minimise the timing residuals by varying the parameters of the timing model.

If the coordinates of the observatory in a barycentric reference system are not precisely known, the conversion from the ToAs to barycentric arrival times will be incorrect and timing residuals will be induced.  A search for such residuals was carried out by \citet{champion2010measuring} who attempted to identify errors in the solar system ephemeris. We show here that it is also possible to use the timing residuals measured for multiple pulsars to determine the observatory coordinates with a precision that depends on the uncertainties of the measured ToAs and the stability of the pulsars. Of course, the observatory need not be on Earth and so this technique provides the possibility for determining the position of a spacecraft with sufficient accuracy to enable such a probe to be navigated in the solar system.

\citet{wallace1988radio} showed that pulsar observations using radio telescopes would require a large antenna which would be impractical for most spacecraft. The possibility of spacecraft navigation using optical observations of pulsars was also shown to be impractical because of the small number of detectable optical pulsars (optical pulsations have only been seen for five pulsars; \citealp{shearer2001implications}) and the required size of the optical telescope to detect the pulsations. Since the 1970s it has been known that a relatively large fraction of pulsars are detectable using X-ray telescopes. \citet{chester1981navigation} first proposed X-ray observations of pulsars for spacecraft navigation with the major advantage over radio observations being that only a relatively small X-ray telescope is required to detect pulsars.


The basic concept for spacecraft navigation using X-ray pulsar timing is straightforward and was set out by \citet{sala2004feasibility,sheikh2005use,sheikh2006recursive,sheikh2006spacecraft}.  Observations of pulse arrival times are made on the spacecraft of one or more pulsars\footnote{Note that \citet{huang2012navigation} discussed the possibility of using the orbital motion of binary pulsars for spacecraft navigation.  Here we only make use of the rotational periodicity.}.  These arrival times are converted to barycentric arrival times using the best estimate of the spacecraft position.  The barycentric arrival times are compared with the predictions from an existing pulsar ephemeris and the resulting residuals minimised by allowing for small errors in the spacecraft position and velocity. However, the details are complex and any algorithm must account for:

\begin{itemize}

\item Unmodelled pulsar spin irregularities.  Long-term observations of pulsars have unmodelled irregularities in the pulsar's pulse rate \citep{hobbs2010analysis} and indicate sudden changes in the pulsar spin period known as a ``glitch'' (e.g., \citealt{espinoza2011study}).  These irregularities cannot be predicted and their effect must be included in the analysis procedure.  

\item Irregularities in the clock to which the ToAs are referenced. The inevitable errors in the spacecraft clock will cause an apparent drift in the measured pulsar ToAs.

\item Pulse counting ambiguities when determining timing residuals. The pulsar timing residuals correspond to the measured barycentric arrival time minus the time of the \emph{closest} predicted pulse arrival time.  If the error in determining the time delay from the spacecraft to the barycentre is greater than half of the pulse period then the residual will be formed using an incorrect prediction of the barycentric arrival time.

\item Different pulse shapes at different observing frequencies.  The spacecraft will measure a ToA by cross-correlating a standard template of the pulsar in the X-ray band with the actual observation \citep{taylor1992pulsar}.  The pulsar timing ephemeris is likely to have been obtained from long-term, ground-based, radio observations of the pulsars.  The pulse shape may significantly differ between the X-ray and radio wave bands leading to an offset between the ToAs measured using the different bands.

\item The time between pulsar observations.  It may be possible for a spacecraft to observe multiple pulsars simultaneously (either with a wide field-of-view telescope or by using multiple telescopes simultaneously).  However, a simpler system would only allow observations of a single pulsar at a given time.  Between observations the spacecraft may have moved a considerable distance.
\end{itemize}

Some of these issues have been discussed in the literature. \citet{sun2010research} proposed a method to mitigate any error in the spacecraft clock, however this method relies on a model of the clock. \citet{liu2011pulsar} described algorithms for pulsar-based navigation based on Kalman filtering, but did not account for all the issues that affect pulsar observations. \citet{qiao2009development} ignored the pulse ambiguity issue by assuming that the error in the spacecraft position was small compared with the distance light travels in half of the pulse period. \citet{mao2009analysis} attempt to deal with the ambiguity problem using a difference-measurement technique; but this technique needs at least five pulsars or the assistance of a known clock offset.

Our algorithms make use of the standard pulsar timing procedures that are based on linear least-squares-fitting algorithms.  This allows us to include our algorithms as part of the \textsc{tempo2} pulsar timing software package.  In Section 2, we present our two algorithms. In Section 3,  we demonstrate how data sets can be simulated in order to test the algorithms. In Section 4, we present our results and discuss non-simultaneous observations, the clock on the spacecraft and the pulsar timing ephemerides. We present concluding remarks and summarise our paper in Section 5.


\section{Algorithms}

\begin{figure}[H]
\centering 
\includegraphics[height=9cm]{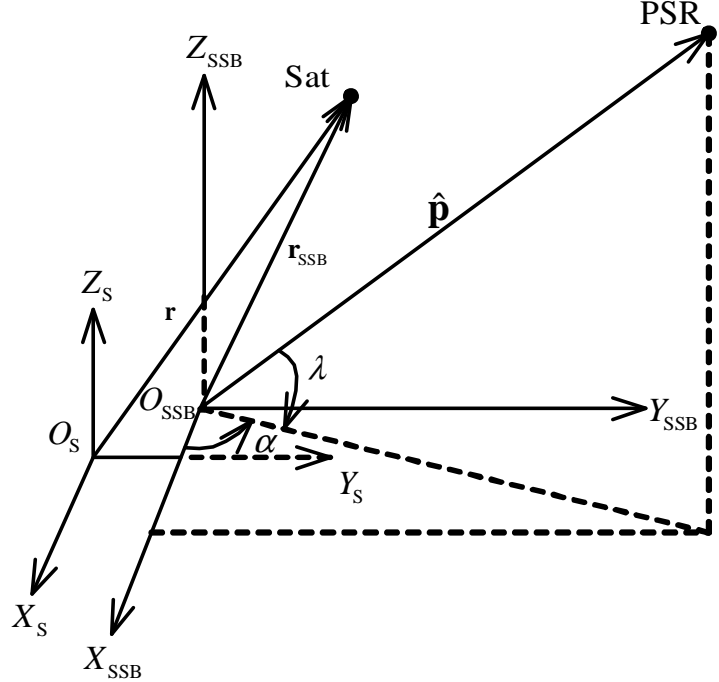}\\ 
\caption{Geometric relationship between the spacecraft, pulsar, Sun and the SSB\label{fg:geoRela}}
\end{figure}

Throughout this paper we assume that a ground-based radio telescope has observed an array of different pulsars for many years before the spacecraft is launched.  We then assume that the spacecraft has an on-board X-ray telescope for observing the same array of pulsars.  The radio data provide 1) timing models for each pulsar and 2) a statistical measure of the noise in the post-fit timing residuals for each pulsar.

Pulsar timing requires a reference frame that is inertial relative to the pulsars.   Following standard pulsar timing procedures we assume that the Solar System Barycentre (SSB) provides a sufficient approximation to an inertial reference frame.  As the main purpose of this work is to determine the coordinates of a spacecraft, we can also use the SSB as our coordinate system.   In Figure~\ref{fg:geoRela} we show the geometric relationship between the spacecraft position (Sat), the pulsar (PSR), the Sun ($O_\mathrm{s}$) and the SSB ($O_\mathrm{SSB}$).  $\alpha$ and $\lambda$ represent the right ascension and declination of the pulsar respectively. ${\bf{r}}_\mathrm{SSB}=[x_\mathrm{SSB}~~y_\mathrm{SSB}~~z_\mathrm{SSB}]^\mathrm{T}$ represents the position of the spacecraft with respect to the SSB, $\bf\hat{p}=\left[\cos{\lambda}\cos{\alpha}~~\cos{\lambda}\sin{\alpha}~\sin{\lambda}\right]^\mathrm{T}$ is a unit vector in the direction of the pulsar\footnote{We assume that the pulsar is at sufficient distance that the angles do not change with the position of spacecraft in the solar system and that the motion of Earth and pulsar can be accounted for by modelling the pulsar proper motion and parallax.}. We also define the velocity of the spacecraft with respect to the SSB as $\dot{{\bf{r}}}_\mathrm{SSB}=\left[v_\mathrm{SSBx}~~v_\mathrm{SSBy}~~v_\mathrm{SSBz}\right]^\mathrm{T}$, and then we can define ${\bf{x}}_\mathrm{SSB}=\left[{\bf{r}}_\mathrm{SSB}^\mathrm{T}~~\dot{{\bf{r}}}_\mathrm{SSB}^\mathrm{T}\right]^\mathrm{T}=[x_\mathrm{SSB}~~y_\mathrm{SSB}~~z_\mathrm{SSB}~~v_\mathrm{SSBx}~~v_\mathrm{SSBy}~~v_\mathrm{SSBz}]^\mathrm{T}$.



Following standard pulsar timing procedures described in \citet{R-Edwards_2006} each ToA (both X-ray and radio) can be converted to Coordinate Barycentric Time (TCB) at the SSB
\begin{equation}
t_\mathrm{SSB}=t_\mathrm{SC}+\frac{1}{c}{\bf\hat{p}}^\mathrm{T}{\bf{r}}_\mathrm{SSB}+\Delta t
\label{eq:predictTOA}
\end{equation}
where $c$ is the vacuum speed of light and $\Delta t$ represents corrections required for converting from the spacecraft clock to a barycentric time scale along with any other extra delays that are necessary. Throughout this paper we assume that the pulsar timing software package, \textsc{tempo2}, can model all non-spacecraft issues (such as the Shapiro delay and, for radio observations, dispersive delays) with sufficient precision and accuracy for our purposes\footnote{We note that \textsc{tempo2} does not assume that the observer is on Earth.}. We justify this assumption by noting that \textsc{tempo2} is expected to account for all known physical effects at the 1\,ns level \citep{G-Hobbs_2006} and has already been used successfully for numerous high time-precision pulsar experiments (e.g., \citealt{manchester2013}).

\textsc{Tempo2} first calculates barycentric arrival times from Equation~\ref{eq:predictTOA} assuming knowledge of the telescope position with respect to the SSB. Predictions for the measured barycentric arrival times are obtained using a simple ``timing model'' for the pulsar that contains its astrometric, pulse and orbital parameters.  Catalogues exist \citep{manchesteraustralia} that contain such timing models. However these timing models can quickly become out-of-date. Any attempt to navigate a spacecraft using X-ray pulsars will have to include an agreement to monitor the same pulsars with radio telescopes prior to the launch and during the mission because the timing models cannot be extrapolated very far without incurring serious errors \citep{deng2012optimal}.

The differences between the predicted barycentric arrival times using the timing model from the ground-station observations and the ``observed'' barycentric arrival times using the X-ray telescope on board the spacecraft are the pulsar timing residuals.  An error in the assumed position of the spacecraft will lead to an error in  ${\bf{r}}_\mathrm{SSB}$, which we denote by $\delta{\bf{r}}_\mathrm{SSB}$.  If the only error is $\delta{\bf{r}}_\mathrm{SSB}$ then the induced timing residual for the observation of pulsar i will be
\begin{equation}
\label{eq:basic}
R_i =\frac{1}{c}{\bf\hat{p_i}}^\mathrm{T}\delta{\bf{r}}_\mathrm{SSB}.
\end{equation}
Therefore a single pulsar observation can be used to identify the error in the observatory position in the direction of that pulsar with a precision of $c\sigma$ where $\sigma$ is the uncertainty of the ToA determination\footnote{This assumes that $\delta{\bf{r}}_\mathrm{SSB}$ is small such that ${\bf\hat{p_i}}^\mathrm{T}\delta{\bf{r}}_\mathrm{SSB}<c P_{\rm fast}/2$, where $P_{\rm fast}$ is the period of the fastest spinning pulsar in the array.}.  Observations of multiple pulsars in different directions on the sky are required to determine all three components of the spacecraft position.

The ground-based observations are assumed to be obtained at radio observing frequencies (typical observations are made in the 20\,cm wavelength band).   In order to produce a precise pulsar timing model the effects of interstellar dispersion need to be accounted for.  Interstellar dispersion is time variable and therefore it is necessary to track changes in the dispersion during the radio observations.  One procedure has been described by \citet{keith2013measurement}, but relies on the radio telescope having a wide bandwidth or observing simultaneously in two different wavelength bands.  We note that the observing frequency for the space-based X-ray observations is sufficiently high that the effects of variations in interstellar dispersion can be ignored. It is, however, essential that the X-ray and radio observations can be aligned.  This is non-trivial as the X-ray and radio pulse shapes can significantly differ.  For some pulsars (e.g., \citealt{rots2004absolute}) this is possible, but for the vast majority of pulsars this is currently not possible with high precision.  We therefore recommend that, after launch, the space-craft and ground-based telescope enter a calibration stage where the position of the spacecraft is known. During this stage the telescopes would simultaneously observe the set of pulsars that will be used during the navigation enabling the alignment to be determined with sufficient precision.

\subsection{Algorithm 1: Absolute position determination}

We start by assuming no knowledge of the trajectory of the spacecraft.  We assume that the X-ray telescope on board the spacecraft has obtained the ToAs for $N_p$ pulsars simultaneously and that the position of the spacecraft is already known to be better than $c P_{\rm fast}/2$. From the simultaneous ToA measurements we wish to know the current position of the spacecraft.

We have updated the \textsc{tempo2} fitting routines to allow $\delta{\bf{r}}_\mathrm{SSB}$ to be determined as part of the standard linear least-squares-fitting procedure using Equation \ref{eq:basic}. To fit multiple pulsars simultaneously, we use the global fitting routines implemented into \textsc{tempo2} by  \citet{champion2010measuring}. Instructions for making use of these updates to \textsc{tempo2} are described in \ref{appendix:navigation}.  This straight-forward procedure therefore directly gives a determination of the three spatial components of $\delta{\bf{r}}_\mathrm{SSB}$ and their corresponding uncertainties.   If more than three pulsars are observed then the problem for determining the space-craft position is over-determined. As the fitting within \textsc{tempo2} is a linear least-squares procedure such an overdetermined system is not a problem and the observations from the various pulsars are correctly weighted as part of the fit.

We assume that the pulsar timing models have been obtained from the ground-station data.  In an ideal situation the spacecraft has available a recent timing model, however, it may be necessary to extrapolate the timing model to the current observation date.  Where necessary, in this paper we carry out such extrapolation using the algorithms developed by \citet{deng2012optimal}. 

\subsection{Algorithm 2: Orbital element determination}

For interplanetary trajectories the position and velocity of the spacecraft can be modelled using orbital mechanics. The details of modelling a spacecraft over long time intervals is complex as the spacecraft is affected by numerous bodies and effects such as solar radiation pressure.  However, over short time intervals we can model the spacecraft trajectory using simple Newtonian equations that account for the major bodies in the Solar System and ignore the extra complexity that arises when attempting to model solar radiation pressure.  Instead of simply making use of a few simultaneous observations as in algorithm 1, it is therefore possible to use a set of ToAs and basic parameterisation of the spacecraft trajectory in order to determine improved orbital parameters and hence obtain the current position of the spacecraft.

Traditionally interplanetary spacecraft trajectories are parameterised in the heliocentric coordinate system. Using the terminology defined in Figure \ref{fg:geoRela} we define the position and velocity of the spacecraft as ${\bf{r}}=\left[x~~y~~z\right]^\mathrm{T}$ and as $\dot{{\bf{r}}}=\left[v_{x}~~v_{y}~~v_{z}\right]^\mathrm{T}$ respectively, and then we can define ${\bf{x}}=\left[{\bf{r}}^\mathrm{T}~~\dot{{\bf{r}}}^\mathrm{T}\right]^\mathrm{T}=[x~~y~~z~~v_x~~v_y~~v_z]^T$. From this definition of $\bf{x}$, the dynamics model $f({\bf{x}},t)=\bf{\dot{x}}$ of the spacecraft can be written as \citep{montenbruck2000satellite}
\begin{equation} 
f({\bf{x}},t) = \left[ \begin{array}{c}
v_{x} \\
v_{y}\\
v_{z} \\
-\frac{{\mu}x}{r^3}+\sum\limits_{k=1}^{9}\mu_k(\frac{x_k-x}{|{\bf{r}}_k-{\bf{r}}|^3}-\frac{x_k}{r_k^3})\\
-\frac{{\mu}y}{r^3}+\sum\limits_{k=1}^{9}\mu_k(\frac{y_k-y}{|{\bf{r}}_k-{\bf{r}}|^3}-\frac{y_k}{r_k^3})\\
-\frac{{\mu}z}{r^3}+\sum\limits_{k=1}^{9}\mu_k(\frac{z_k-z}{|{\bf{r}}_k-{\bf{r}}|^3}-\frac{z_k}{r_k^3})\end{array} \right]
\label{eq:dynamicsModel}
\end{equation}
where $\mu$ is the gravitational constant of the Sun ($\mu=GM_\odot$ where $G$ is Newton's gravitational constant and M$_\odot$ is the solar mass), $\mu_k$ is the gravitational constant of other major objects in the solar system (we include the eight planets and Pluto), ${\bf{r}}_k=[x_k~~y_k~~z_k]$ is the position of these masses in the heliocentric coordinate system, $r=\sqrt{x^2+y^2+z^2}$ is the distance between the probe and the centre of the Sun, $r_k=\sqrt{x_k^2+y_k^2+z_k^2}$ is the distance between the planets and the Sun. 

Given a set of initial conditions (${\bf{x}_0}$) we numerically integrate Equation~\ref{eq:dynamicsModel} to obtain the estimated ${\bf{x}_j}$ of the spacecraft at the time of each pulsar observation. We form barycentric arrival times and subsequently determine the pulsar timing residuals from the measured spacecraft arrival times.  

We define the error in the initial condition as $\delta{\bf{x}_0}=\left[{\delta\bf{r}_0}^\mathrm{T}~~\delta\dot{{\bf{r}}}_0^\mathrm{T}\right]^\mathrm{T}$. As 1) the integration time intervals are short and 2) we assume that the error in the initial condition is small, the position error at the time of each pulsar observation can be written as $\delta{\bf{r}_j} = {\delta\bf{r}_0} + \delta\dot{{\bf{r}}}_0{t_{j_0}}$, where $t_{j_0}=t_j-t_0$. As the vector error remains unchanged in both coordinate systems, the position error at the time of each pulsar observation in the barycentric reference system will be 
\begin{equation}
\begin{split}
\delta{\bf{r}_\mathrm{SSBj}} &= {\delta\bf{r}_\mathrm{SSB_0}} + \delta\dot{{\bf{r}}}_\mathrm{SSB_0}{t_{j_0}}
\end{split}
\label{eq:two-basic}
\end{equation}
where $\delta{{\bf{x}}_\mathrm{SSB_0}}=[\delta\bf{r}_\mathrm{SSB_0}^\mathrm{T}~~\delta\dot{{\bf{r}}}_\mathrm{SSB_0}^\mathrm{T}]^\mathrm{T}$, $\delta\bf{r}_\mathrm{SSB_0}$ and $\delta\dot{{\bf{r}}}_\mathrm{SSB_0}$ are the errors of the initial position and velocity in the SSB respectively. From Equation \ref{eq:two-basic}, we can update Equation \ref{eq:basic} to

\begin{equation}
{R_\mathrm{ij}}=\frac{1}{c}[{\bf\hat{p}}_\mathrm{i}^T~~{\bf\hat{p}}_\mathrm{i}^T{t_\mathrm{j_0}}]\delta{{\bf{x}}_\mathrm{SSB_0}}.\\
\label{eq:Rel-final}
\end{equation}

Using Equation \ref{eq:Rel-final}, we can therefore carry out a standard linear least-squares-fit procedure using multiple pulsars to fit for the error in the initial parameters of the trajectory. We have implemented this algorithm into \textsc{tempo2} as the \textsc{navOrbit} plugin. It allows the user to determine ${{\bf{x}}_\mathrm{SSB_0}}$ as part of the standard linear least-squares-fitting procedure.  The \textsc{navOrbit} plugin implements the following algorithm (usage details are given in \ref{appendix:navigation}:
\begin{itemize}
\item A set of pulsar timing models are loaded for each pulsar
\item A set of arrival times are loaded for each pulsar
\item The assumed position of the spacecraft is converted to heliocentric coordinates
\item For a given time the positions of the planets and moon are determined in heliocentric coordinates
\item Equation~\ref{eq:dynamicsModel} is numerically integrated to estimate the position and velocity of the spacecraft at the time of each pulsar observation
\item The linear least-squares-fitting algorithm is used to determine the error in the initial estimates of the spacecraft position and velocity by fitting to each pulsar data set simultaneously
\item These errors in the initial estimates are used to produce the final output that gives the best estimate of the spacecraft's initial position and velocity.
\end{itemize}

\subsection{Summary of method}

\begin{figure*}
\begin{center}\includegraphics[width=13cm]{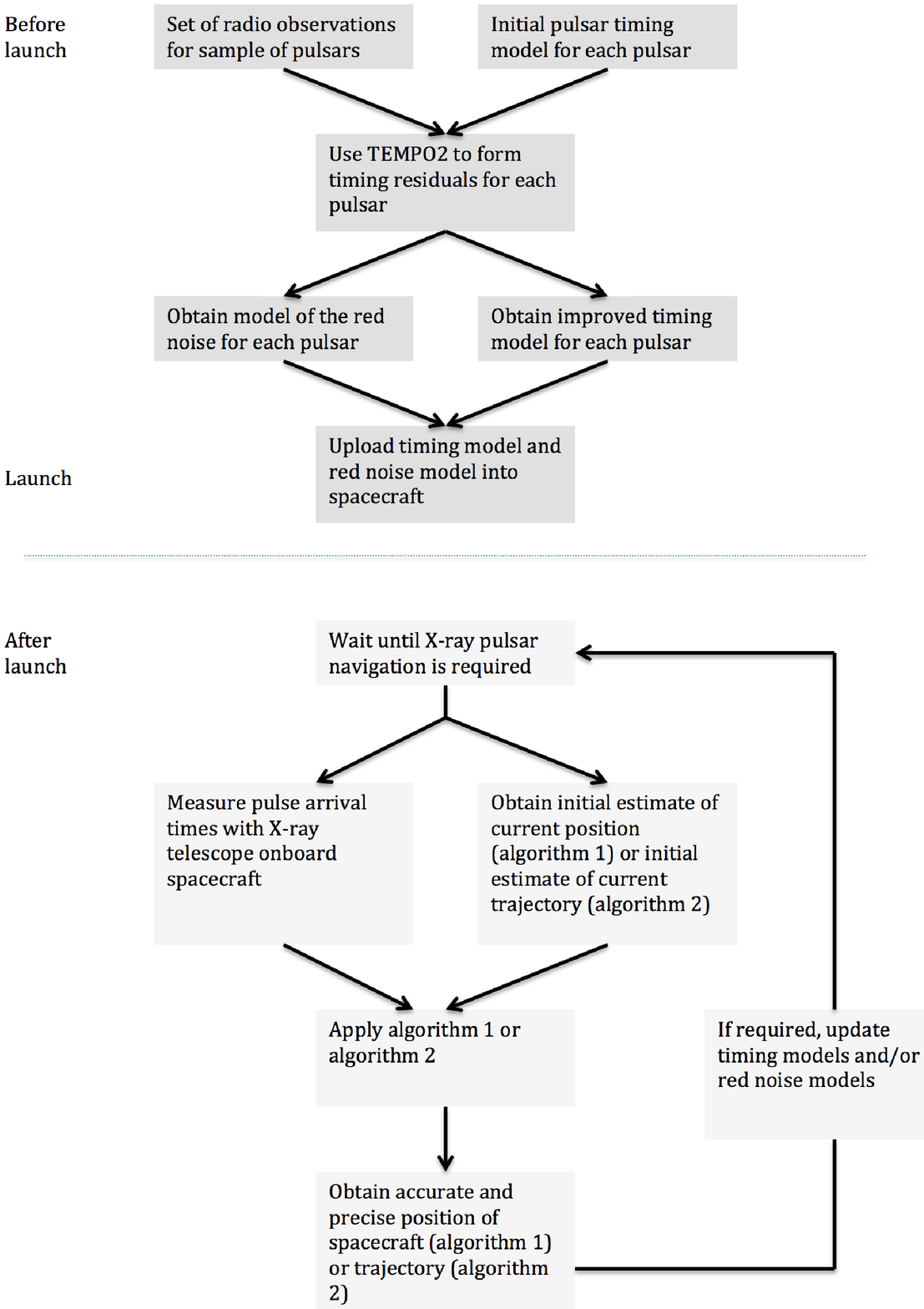}\end{center}
\caption{Summary of the procedures described in this paper for spacecraft navigation.  This flow-chart described the observations and processing carried out at the ground-station prior to launch (top), 2) that these results are loaded into the spacecraft before launch and 3) the basic procedures followed on-board the spacecraft in order to determine its position.  \label{fg:flowchart1}}
\end{figure*}

\begin{figure*}
\begin{center}\includegraphics[width=13cm]{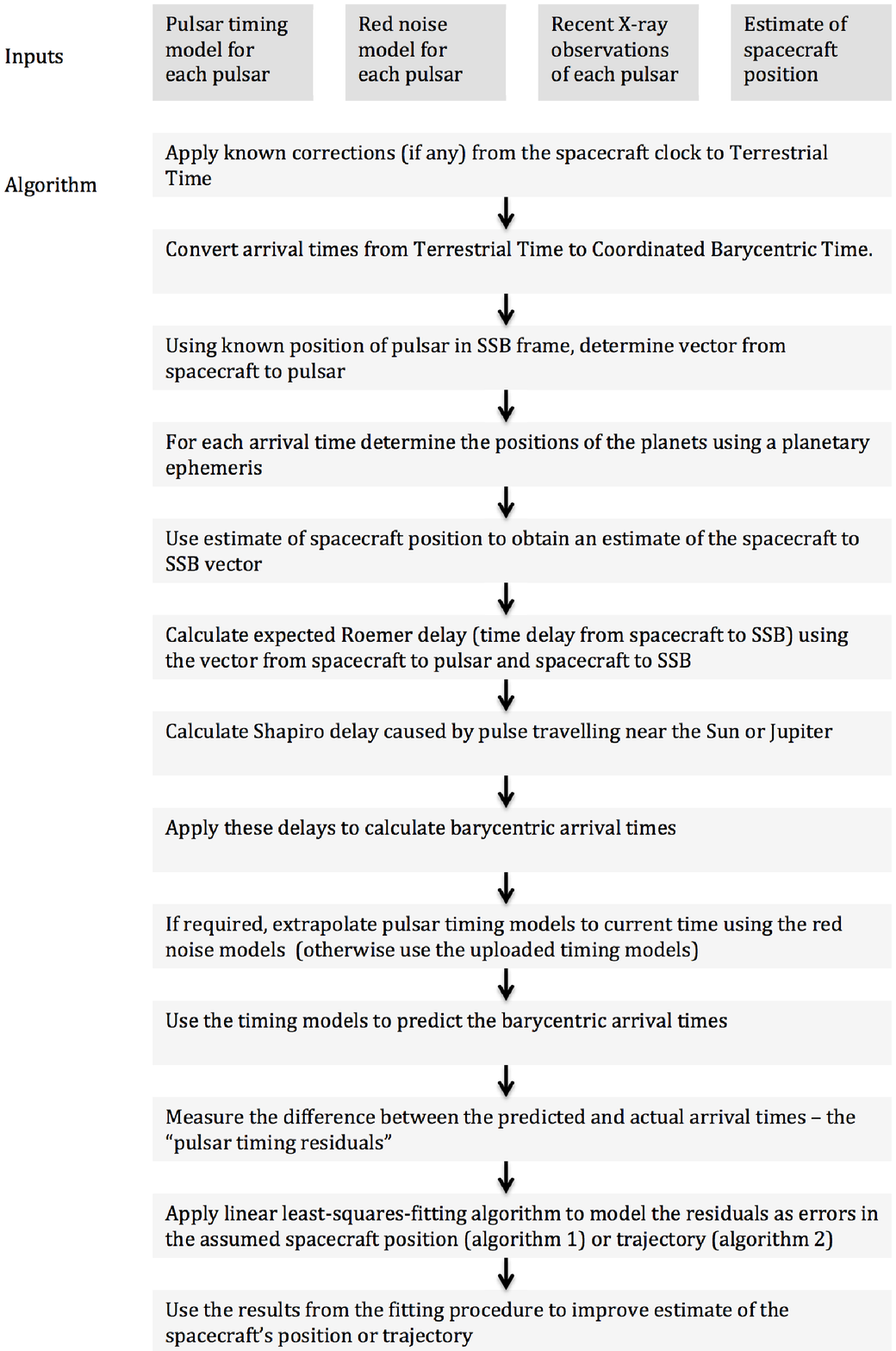}\end{center}
\caption{Summary of the procedures described in this paper for spacecraft navigation.    This flow-chard provides an overview of how \textsc{tempo2} processes pulse arrival times in order to obtain the spacecraft position.  Details of algorithm 1 and algorithm 2 are provided in this paper.  Details of the various \textsc{tempo2} algorithms (such as determining the Shapiro delay etc.) are provided in \citet{R-Edwards_2006}.\label{fg:flowchart2}}
\end{figure*}

An overview of our method and of the necessary data processing is shown in Figures~\ref{fg:flowchart1} and \ref{fg:flowchart2} and our method can be summarised as follows. Before the spacecraft is launched:

\begin{itemize}
\item A ground station radio telescope has been observing a set of pulsars that can be observed in both the radio and X-ray observing bands over time spans of years.  In our paper we assume that 4 pulsars are being observed for approximately 5\,yr.
\item These ground station data sets have been used to obtain a model of the spectrum of the low-frequency noise present in the timing residuals for each pulsar.
\item These ground station data sets are also used to produce a model for the pulsars' pulse, astrometric and binary parameters (the pulsar ``timing models").
\end{itemize}

For most of this paper, we assume that the pulsar timing model and the model for the low-frequency noise present is uploaded into the spacecraft computers before launch. After the launch of the spacecraft we consider two questions.  The first is ``what is the current position of the spacecraft without any knowledge of the spacecraft trajectory?" and the second is "what is the position and velocity of the spacecraft at a specified time assuming that the spacecraft is in a simple trajectory?".  For autonomous navigation we assume that this has to be calculated on board the spacecraft.  At the time that the spacecraft requires a position estimate:

\begin{itemize}
\item An X-ray telescope on-board the spacecraft measures the arrival times of the same sample of pulsars that have been observed at the ground station.  
\item An approximate position (and, if necessary, an approximate velocity) of the spacecraft is obtained. This could be from previous estimates, or from other less-precise navigational aids (see Section 5).
\item If the arrival times for multiple pulsars have been measured simultaneously then the onboard computer applies either algorithm 1 or algorithm 2 (the choice is discussed below).  The output from the algorithm indicates the error in the estimated spacecraft position (with its corresponding uncertainty).  The onboard computer can then update the estimate of its current position.  Algorithm 2 also provides an estimate of the spacecraft velocity.
\item If the arrival times are not simultaneous then the onboard computer applies algorithm 2. This provides an update to the current estimation of the trajectory of the spacecraft. 
\item The procedure is repeated as necessary for the onboard computer to keep track of the spacecraft's position (and, for algorithm 2 velocity).
\end{itemize}

\section{Data set}

In order to demonstrate the effectiveness of our methods we have simulated 1) the trajectory of a spacecraft leaving an Earth orbit  and travelling to Mars, 2) the ToAs measured using X-ray telescope on the spacecraft and 3) the ToAs measured using a radio telescope at the ground station.

\subsection{Choice of pulsars}

\begin{table*}[!t]
\centering
\begin{footnotesize}
\caption{Selected rotation-powered pulsars. The columns contain each pulsar's period (P), right-ascension (R.A.) and declination (Dec.), coordinate of the pulsar in the barycentric reference frame and information relating to the red noise detected in the residuals. The red noise spectral density is defined with a representative power, $P_0$, spectral exponent, $\alpha$ and corner frequency, $f_c$. \label{Tab:Pulsar-list}}
 
\begin{tabular}{llccccccccc} \hline
Jname			&Bname		&$P$		&R.A.		&Dec. &${\bf\hat{p}}_x$ &${\bf\hat{p}}_y$ &${\bf\hat{p}}_z$		& $P_0$ 		& $\alpha$ 	& $f_c$\\ 
			&		&(ms)		&($^\circ$)	&($^\circ$) 	& & & 	& (y$^3$) 		& 		& (y$^{-1}$) \\ \hline							
J0437$-$4715		&		&5.75		&69.32		&$-$46.75 	&0.24 	&0.64 &$-0.73$	&$1.4\times10^{-29}$ 	&4 &0.8	\\
J1824$-$2452A		&B1821$-$24A	&3.05		&276.13		&$-$23.13 	&0.10 	&$-0.90$ &$-0.42$ 	&$1.1\times10^{-24}$ 	&3 &0.1	\\
J1939$+$2134		&B1937$+$21	&1.55		&294.91		&$+$21.58 	&0.39 	&$-$0.84 &0.37		&$1.0\times10^{-25}$ 	&4 &0.1	\\
J2124$-$3358		&		&4.93		&321.18		&$-$32.02 	&0.65	 &$-$0.52 &$-0.56$	&-		     	&-	&-\\\hline
\end{tabular}
\end{footnotesize}
\end{table*}

More than 2000 pulsars are now included in the Australia Telescope National Facility (ATNF) pulsar catalogue \citep{manchesteraustralia}, but not all of these pulsars are suitable for pulsar-navigation experiments.  We have selected a group of four millisecond pulsars that have been observed in both radio and X-ray bands. These pulsars are relatively faint, but are extremely stable in their rotation and can be timed (with a large radio telescope) with microsecond timing precision. 

In order to obtain realistic predictions of the precision with which the ToAs could be measured and of the rotational irregularities that will affect the long-term stability of these pulsars, we have selected pulsars that have been observed by the 64-m Parkes radio telescope over many years. These pulsars have been observed as part of the Parkes Pulsar Timing Array (PPTA) project (\citealt{manchester2013}). X-ray observations for PSRs~J0437$-$4715 and J2124$-$3358 can be found in \citet{becker1998x}. X-ray observations of PSR~J1824$-$2452A have been published by \citet{rutledge2004microsecond} and observations of PSR~J1939$+$2134 by \citet{cusumano2003phase}. In the first five columns of Table~\ref{Tab:Pulsar-list} we list (from the ATNF pulsar catalogue; \citealt{manchesteraustralia}) the J2000 and B1950 names for each pulsar, the pulse period ($P$) and the right ascension and declination of the pulsars.  The post-fit timing residuals for these pulsars are shown in the left-hand panels of Figure~\ref{Fg:dset1}.

These timing residuals show unexplained low-frequency noise (particularly PSRs J1824$-$2452A and J1939$+$2134), which we subsequently refer to as ``red timing noise'' (described in more detail in Section \ref{sect:updatingephemerides}).  For the first three pulsars, we model the power spectral density  of the red timing noise as
\begin{equation}\label{eq:spectralModel}
P(f) = \frac{P_0}{\left(1+[f/f_c]^2\right)^{\alpha/2}}
\end{equation}
using the \textsc{spectralModel} plugin to the \textsc{tempo2} software package (see e.g., \citealt{W-Coles_2011}). Where ${P_0}$ is a measure of the strength of the red timing noise, $f_c$ is a corner frequency that ensures that the red noise model turns over at low frequencies and $\alpha$ is the spectral exponent of the red noise. Instructions for using this plugin are given in \ref{appendix:data_analysis}. The values of $P_0$, $f_c$ and $\alpha$ for each pulsar are given in the last three columns of Table~\ref{Tab:Pulsar-list}.

\subsection{Simulated ground-station data}
\label{sect:ground-data}
\begin{figure*}[!t]
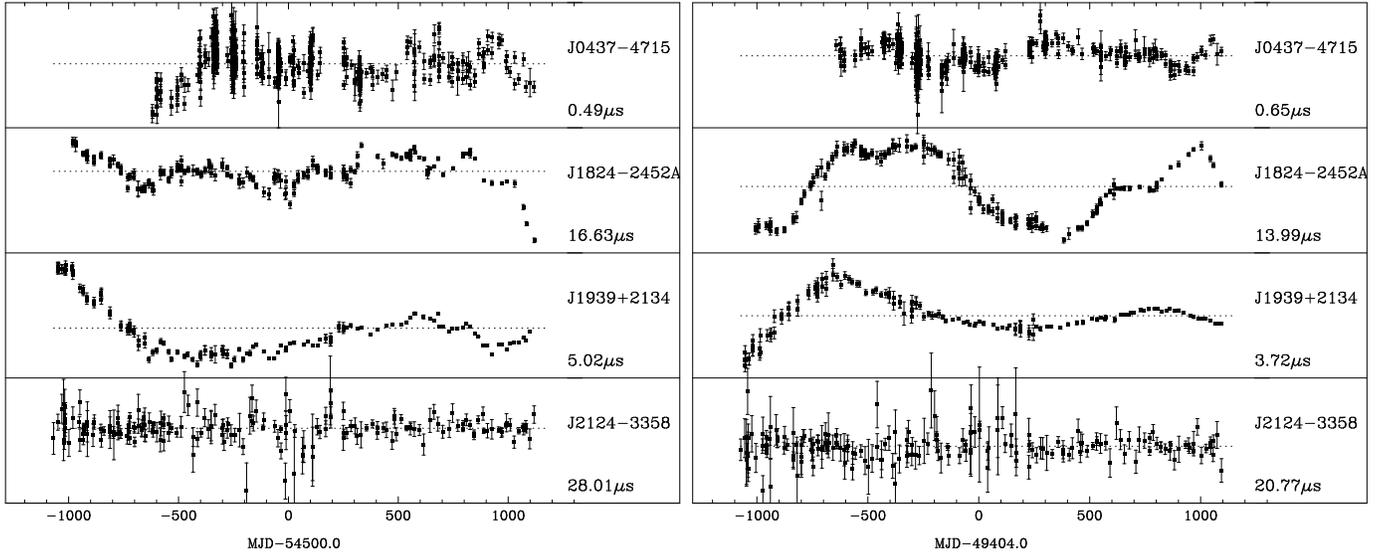

\centering
\includegraphics[height=9cm,angle=-90]{dataset1_real.ps}
\includegraphics[height=9cm,angle=-90]{dataset1.ps}
\caption{Timing residuals for ground-station observations of the millisecond pulsars array. The figure on the left shows the real timing residuals. The figure on the right shows one realisation of simulated timing residuals. The text on the right of each plot gives the pulsar's name and the value gives the rms timing residual.\label{Fg:dset1}} 
\end{figure*}

We assume, for the purposes of this simulation, that the observing cadence and ToA precision are the same as the actual observations from the PPTA project. Our ground-station data set is therefore formed by shifting the actual Parkes observations in time so that the last observation occurs 10 days before the simulated launch\footnote{We define the ``launch'' as the beginning of the simulated trajectory, which will be described in Section \ref{Sect:orbit-simulation}.} of the spacecraft. The observing cadence for each pulsar and the white noise level for each ToA remain as in the actual Parkes observations. ``Idealised site-arrival-times'' are formed to produce a set of arrival times that perfectly match the input timing model using the \textsc{formIdeal} plugin to \textsc{tempo2}. Details for forming idealised site-arrival-times are given in \citealt{hobbs2010analysis}.  In brief, a set of trial site-arrival-times are processed within \textsc{tempo2} to form timing residuals with respect to the given timing model.  Each residual is then subtracted from the corresponding site-arrival-times and the process iterated until the timing residuals are consistent with zero residual.  We note that this procedure accounts for all the physical phenomena that has been included within the \textsc{tempo2} software package, such as the Roemer and Shapiro delays. Gaussian white noise (that represents the uncertainty in measuring the ToAs) is added to the ideal site-arrival-times using the \textsc{addGaussian} plugin.  The red timing noise is included using the measured spectral parameters of the red noise using the \textsc{addRedNoise} plugin. Usage instructions for these plugins can be found in \ref{appendix:data_simulation}. One realisation of the resulting simulated timing residuals are shown in the right-hand panel of Figure~\ref{Fg:dset1}.  We note that the red noise level compared to the white noise level is similar between the actual data and the simulation and therefore these simulations provide a reasonable estimate of the noise that we may expect in the timing residuals for such pulsars.

\subsection{Simulated trajectory from Earth to Mars}
\label{Sect:orbit-simulation}

\begin{table*}[!t]
\centering 
\begin{footnotesize}
\caption{The initial parameters for the trajectory from Earth to Mars in terms of the Keplerian parameters\label{Tab:orbit-parameters}}
\begin{tabular}{ll} \hline
Parameter 				&Value\\\hline
Orbit Epoch (UTC)			&1997 Mar 01 00:00:00.000\\
Semi-major axis (km)			&193216365.381\\
Eccentricity				&0.236386\\
Inclination ($^\circ$)			&23.455\\
Right ascension of Asc. Node ($^\circ$)	&0.258\\	
Argument of Perihelion ($^\circ$)	&71.347\\
True Anomaly ($^\circ$) 		&85.152\\\hline 		
\end{tabular}
\end{footnotesize}
\end{table*}

\begin{table*}[!t]
\centering 
\begin{footnotesize}
\caption{The initial parameters for the trajectory from Earth to Mars in Cartesian coordinates\label{Tab:orbit-state}}
\begin{tabular}{ll} \hline
Parameter 		&Value\\\hline
Orbit Epoch (UTC)	&1997 Mar 01 00:00:00.000\\
$x$ (km)		&$-$164304979.5\\
$y$ (km)		&64685884.9\\
$z$ (km)		&28386557.2\\
$v_x$ (km/s)		&$-$16.7\\
$v_y$ (km/s)		&$-$20.9\\
$v_z$ (km/s)		&$-$9.0\\\hline 		
\end{tabular}
\end{footnotesize}
\end{table*}

\begin{figure*}[!t]
\centering
\begin{minipage}{9cm}\includegraphics[height=7.5cm, angle = 0]{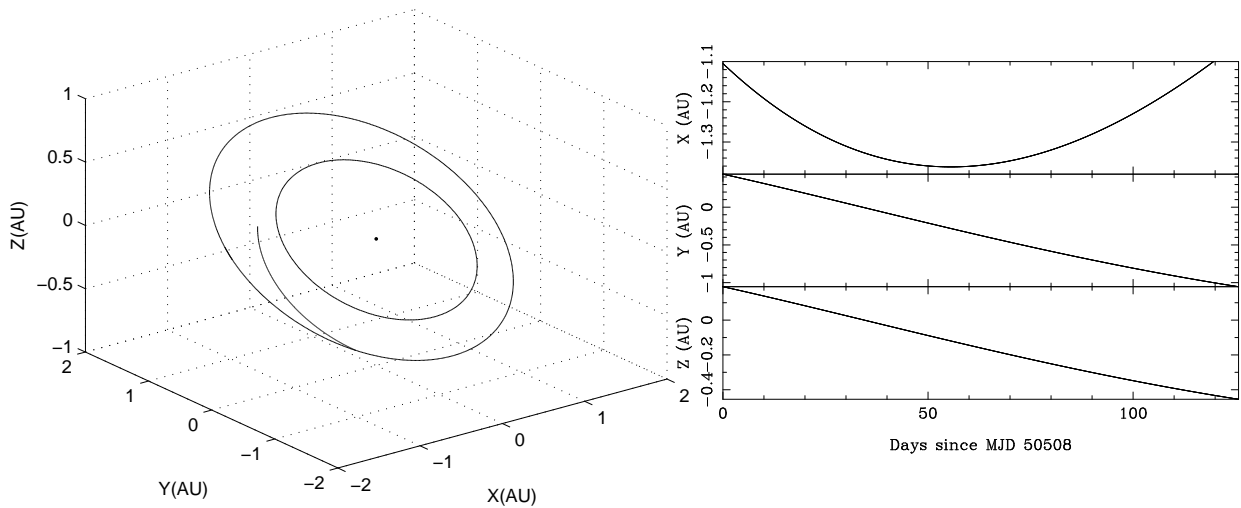} \end{minipage}
\begin{minipage}{9cm}\includegraphics[height=7.5cm, angle=-90]{Earth-to-Mars-orbit2.ps}\end{minipage}
\caption{The simulated trajectory from Earth to Mars. The left figure shows the SSB (the point at the centre), the orbit of Earth (the inner circle), the orbit of Mars (the outer circle) and the simulated trajectory of spacecraft (the line between two circles). The panels on the right show the x, y and z coordinates of the spacecraft versus time.  \label{Fg:orbit-simulation}} 
\end{figure*}

In order to simulate the trajectory of a spacecraft travelling from Earth to Mars, we have used the \textsc{Astrogator} part of the \textsc{STK} software\footnote{{https://www.agi.com/products/by-product-type/applications/stk/add-on-modules/stk-astrogator/}}. The parameters for the trajectory are chosen to simulate the cruise phase of NASA's Pathfinder mission. We take the orbit epoch and Keplerian parameters from \citet{Mars_Pathfinder}, which are listed in Table \ref{Tab:orbit-parameters}. The simulated trajectory (shown in Figure \ref{Fg:orbit-simulation}) takes 4 months and ends at the closest point to Mars. The initial Keplerian parameters have been converted to the cartesian coordinate system, which are shown in Table \ref{Tab:orbit-state}. In the simulation, we take the gravitational force of the Sun and other major objects (including the eight planets and Pluto) and the solar radiation pressure into account. 

In order to account for solar radiation pressure we assume that our spacecraft has similar properties to the Pathfinder mission spacecraft \citep{Mars_Pathfinder}. The Pathfinder weighed 895 kg including 94 kg of propellant at launch, measured 2.65 meters in diameter, was 1.5\,m tall and powered by 2.5\,m$^2$ gallium arsenide solar cells \citep{Mars_Pathfinder}.  In order to simplify the calculations, we assume that 1) the weight of the spacecraft remains unchanged during the whole trajectory, 2) the solar radiation pressure coefficient can be derived from \citet{1991_Gordon} and 3) the solar radiation pressure can be modelled as spherical radiation pressure and that the solar radiation pressure area is equal to the area of solar cells plus the maximum cross-sectional area of the spacecraft\footnote{We can make these assumptions as details of the solar radiation pressure does not significantly affect the results of this paper. More realistic solar radiation pressure model and solar radiation pressure coefficient can be found in numerous papers (e.g., \citealp{kubo1999solar}).}. With these assumption, we set 1) the dry mass of the spacecraft as 895 kg and the fuel mass as 0 kg, 2) the solar radiation pressure coefficient as 1.25 and 3) the solar radiation pressure area as 8.0155\,m$^2$ in the \textsc{STK} software. We keep other parameters for the trajectory simulation as default. 

\subsection{Simulated X-ray data}
The ToAs measured by the X-ray telescope on board the spacecraft are simulated by 1) forming idealised site-arrival-times with respect to the nominal pulsar ephemeris every seven days during the mission, 2) adding Gaussian white noise according to the predicted ToA uncertainties (see below) and 3) adding red timing noise according to the spectral model for each pulsar that was determined using the radio observations\footnote{This assumes that the red timing noise is intrinsic to the pulsar and not caused by instrumental issues or delays as the pulse propagates through the interstellar medium \citep{rutledge2004microsecond} and that the dominant effects of the interstellar medium have been removed in the Parkes data sets \citep{keith2013measurement}.}.

For some of these pulsars it would be possible to use X-ray ToA uncertainties that have been measured (e.g., the X-ray ToA uncertainties of PSR~B1821$-$24 from \citet{rutledge2004microsecond} are about 10 $\mu$s.).  However, only a few X-ray observations have been made and various different telescopes used.  For our simulations we predict the expected X-ray ToA uncertainties according to the signal-to-noise ratio of the pulse profile. The typical signal-to-noise ratio of an X-ray pulse profile can be estimated from \citep{sheikh2005use}

\begin{equation}
\mathrm{SNR}=\frac{F_XAp_ft_{\mathrm{obs}}}{\sqrt{[B_X+F_X(1-p_f)](At_{\mathrm{obs}}d)+F_XAp_ft_{\mathrm{obs}}}}\label{eq:SNR}
\end{equation} 
where $F_X$ is the observed X-ray photon flux from the pulsar, $p_f$ is the pulsed fraction, $B_X$ is the X-ray background radiation and $A$ is the area of the detector. The duty cycle $d=W/P$ of a pulse is the pulse width $W$ as a fraction of the pulse period $P$.   For this work we assume that the X-ray detector on board spacecraft has design parameters similar to the Rossi X-ray Timing Explorer (RXTE).  We therefore assume an effective area of $\sim$\,6500\,cm$^2$  and the background $B_X$ to be 2.0\,mCrab \footnote{\url{http://heasarc.gsfc.nasa.gov/docs/xte/PCA.html}} $\approx$\,4.40$\times 10^{-11}$\,erg{s}$^{-1}${cm}$^{-2}$ \footnote{According to \citet{kirsch2005crab}, 1.0\,mCrab =2.20$\times10^{-11}$\,erg{s}$^{-1}${cm}$^{-2}$ for the 2-10\,keV energy band.}. Using the conversion equation described in \citet{sheikh2005use}, the background is equal to 2.2\,$\times10^{-2}$\,ph{s}$^{-1}${cm}$^{-2}$. We assume a typical observation duration of $t_{\mathrm{obs}}$=3600\,s and that the on-board X-ray telescope can be pointed with sufficient accuracy to detect the pulsar. We note that the Proportional Counter Array (PCA) has a 1 degree FWHM field of view and therefore any telescope used for navigation will require a similar pointing accuracy. Approximate ToA uncertainties can subsequently be obtained from:

\begin{equation}
\sigma_{ToA}=\frac{1}{2}W/\mathrm{SNR}\label{eq:sigma_toa}.
\end{equation}

We get the flux density $F_X$ (2-10\,keV), pulse width $W$ and the pulsed fraction $p_f$ for PSRs J0437$-$4715, J1824$-$2452A, J1939$+$2134, J2124$-$3358 from \citet{sheikh2005use}. All of these parameters are summarized in Table \ref{Tab:error-bar}.  We note that, in practice, the ToA uncertainties obtained from the X-ray observations may be more or less precise than the values given in this table.  For instance, the pulse profiles can have multiple components and so $W$ in Equation~\ref{eq:sigma_toa} is not well defined.  Of course, it would not possible to determine a ToA for PSR J2124$-$3358 if the signal-to-noise ratio was $<$1 as indicated using the above equations.  In that case the integration time would need to be increased in order to obtain a ToA for that specific pulsar. However, due to the uncertainties in the possible ToA precision we have, for this paper, assumed that a ToA uncertainty of $\sim 0.2$\,ms (as listed in the table) would be possible for this pulsar.


\begin{table*}[!t]
\caption{Emission properties and ToA uncertainties for X-ray observations of the selected pulsars\label{Tab:error-bar}}
\centering
\begin{footnotesize}
\begin{tabular}{llccccc} \hline
Jname		& Bname  		&$F_X$			&$p_f$		&$W$	&SNR	&$\sigma_{{ToA}}$	\\
		& 			&(ph{s}$^{-1}${cm}$^{-2}$)&		&(ms)	& 	&(s)			\\ \hline						
J0437$-$4715	& 			&6.65$\times10^{-5}$ 	&0.275		&0.290	&2.61 	&5.6$\times10^{-5}$	\\
J1824$-$2452A	& B1821$-$24A		&1.93$\times10^{-4}$ 	&0.980		&0.055	&37.51 	&7.3$\times10^{-7}$	\\
J1939$+$2134	& B1937+21 		&4.99$\times10^{-5}$ 	&0.860		&0.021	&11.15 	&9.4$\times10^{-7}$	\\
J2124$-$3358	& 			&1.28$\times10^{-5}$ 	&0.282		&0.250	&0.52 	&2.4$\times10^{-4}$	\\
\hline
\end{tabular}
\end{footnotesize}
\end{table*}

\section{Results and discussion}

\begin{table*}[!t]
\centering
\begin{footnotesize}
\caption{Summary of the simulations in the paper\label{Tab:basic_summary}}
\begin{tabular}{lcccccccccccc} \hline
Test	&Alg.   & Scenario	&Extrap. 	&Non-sim. 	&Clk err.	&Clk corr.	&Glitch		&Radio		&X-ray &Auto.\\\hline
1	&1	& I &$\times$ 	&$\times$ 	&$\times$	&$\times$ 	&$\times$ 	&$\times$	&$\times$       &$\surd$\\
2	&1	& I &$\surd$	&$\times$ 	&$\times$	&$\times$ 	&$\times$ 	&$\times$	&$\times$       &$\surd$\\
3	&2	& II &$\times$ 	&$\times$ 	&$\times$	&$\times$ 	&$\times$ 	&$\times$	&$\times$       &$\surd$\\
4	&2	& II &$\surd$	&$\times$ 	&$\times$	&$\times$ 	&$\times$ 	&$\times$ 	&$\times$       &$\surd$\\
5	&2	& III &$\surd$	&$\surd$ 	&$\times$	&$\times$ 	&$\times$ 	&$\times$ 	&$\times$       &$\surd$\\\\
6	&2	& IV &$\surd$ 	&$\times$ 	&$\surd$ 	&$\times$ 	&$\times$ 	&$\times$ 	&$\times$       &$\surd$\\
7	&2	& IV &$\surd$ 	&$\times$ 	&$\surd$ 	&$\surd$ 	&$\times$ 	&$\times$ 	&$\times$       &$\surd$\\\\
8	&1	& V &$\surd$ 	&$\times$ 	&$\times$ 	&$\times$ 	&$\surd$ 	&$\times$ 	&$\times$       &$\surd$\\
9	&1	& V &$\surd$ 	&$\times$ 	&$\times$ 	&$\times$	&$\surd$ 	&$\surd$ 	&$\times$       &$\times$\\
10	&1	& V &$\surd$ 	&$\times$ 	&$\times$ 	&$\times$ 	&$\surd$ 	&$\times$ 	&$\surd$        &$\surd$\\\\

11      &1      & VI &$\surd$	&$\times$ 	&$\times$	&$\times$ 	&$\times$ 	&$\times$	&$\times$       &$\times$ \\
\hline
\end{tabular}
\end{footnotesize}
\end{table*}

\begin{table*}[!t]
\centering
\caption{Mean of the error bar in the position determination on arrival at Mars for the 100 realisations of the simulated data sets\label{Tab:errorbar_result}}
\begin{footnotesize}
\begin{tabular}{lccccccccccccccc} \hline
Test 	&Alg.   &$\sigma{x_e}$ 	&$\sigma{y_e}$ 	&$\sigma{z_e}$ &$\sigma{v_{xe}}$ 	&$\sigma{v_{ye}}$ 	&$\sigma{v_{ze}}$ 	&$\sigma{clk_e}$ 	\\
	&       &(km)	&(km)	&(km) 	&(kms$^{-1}$)	&(kms$^{-1}$)	&(kms$^{-1}$) &(s)\\\hline
1	&1      &94.9	&27.1 	&37.6 	&- 	&- 	&- 	&-\\
2	&1      &22.0 	&6.4 	&8.7 	&- 	&- 	&- 	&-\\
3	&2      &37.0 	&10.8 	&14.6 	&$7.7\times10^{-4}$ 	&$2.2\times10^{-4}$ 	&$3.0\times10^{-4}$ 	&-\\
4	&2      &8.8 	&2.6 	&3.5 	&$1.8\times10^{-4}$ 	&$5.3\times10^{-5}$ 	&$7.2\times10^{-5}$ 	&-\\
5	&2      &16.1 	&4.7 	&6.4	&$3.7\times10^{-4}$ 	&$1.1\times10^{-4}$ 	&$1.4\times10^{-4}$ 	&-\\\\

6 	&2      &8.9 	&2.6 	&3.5 	&$1.8\times10^{-4}$ 	&$5.4\times10^{-5}$ 	&$7.3\times10^{-5}$ 	&-\\
7 	&2      &27.9 	&5.5 	&10.2 	&$1.8\times10^{-4}$ 	&$5.3\times10^{-5}$ 	&$7.2\times10^{-5}$ 	&$3.7\times10^{-5}$\\\\

8 	&1      &22.1 	&6.4 	&8.7 	&- 	&- 	&- 	&-\\
9 	&1      &22.9 	&6.7 	&9.1 	&- 	&- 	&- 	&-\\
10	&1      &22.0 	&6.4 	&8.7 	&- 	&- 	&- 	&-\\\\

11     &1      &21.9 	&6.4 	&8.7 	&- 	&- 	&- 	&-\\\hline
\end{tabular}
\end{footnotesize}
\end{table*}

\begin{table*}[!t]
\centering
\begin{footnotesize}
\caption{The rms of the discrepancy on arrival at Mars for the 100 realisations of the simulated data sets\label{Tab:discrepancy_result}}
\begin{tabular}{lccccccccc} \hline
Test   &Alg. 	&RMS 	&RMS 	&RMS 	&RMS 	&RMS 	&RMS 	&RMS \\	
	&       &$\bigtriangleup{x_e}$ 	&$\bigtriangleup{y_e}$ 	&$\bigtriangleup{z_e}$ &$\bigtriangleup{v_{xe}}$ 	&$\bigtriangleup{v_{ye}}$ 	&$\bigtriangleup{v_{ze}}$ 	&$\bigtriangleup{clk_e}$ \\
	&	&(km)	&(km)	&(km) &(kms$^{-1}$)	&(kms$^{-1}$)	&(kms$^{-1}$) 	&(s) \\\hline
1	&1      &21.7 	&6.4 	&8.5	&- 	&- 	&- 	\\
2	&1      &24.6 	&7.2 	&9.8 	&- 	&- 	&- 	\\
3	&2      &10.1 	&3.0 	&4.1 	&$2.2\times10^{-4}$ 	&$6.5\times10^{-5}$ 	&$8.81\times10^{-5}$ 	&- \\
4	&2      &9.4 	&2.8 	&3.8 	&$1.9\times10^{-4}$ 	&$5.4\times10^{-5}$ 	&$7.3\times10^{-5}$ 	&- \\
5	&2      &16.2 	&4.7 	&6.5 	&$3.6\times10^{-4}$ 	&$1.0\times10^{-4}$ 	&$1.4\times10^{-4}$ 	&- \\\\

6 	&2      &94.7	&17.5 	&34.1	&$1.77\times10^{-4}$ 	&$5.1\times10^{-5}$ 	&$7.0\times10^{-5}$ 	&- \\
7 	&2      &26.8	&5.2	&9.9 	&$2.0\times10^{-4}$ 	&$5.6\times10^{-5}$ 	&$7.9\times10^{-5}$ 	&$3.4\times10^{-5}$\\\\

8 	&1      &31.2 	&22.2 	&31.3	&- 	&- 	&- 	&- \\
9 	&1      &23.3 	&6.8 	&9.3	&- 	&- 	&- 	&- \\
10	&1      &21.7 	&6.4 	&8.5	&- 	&- 	&- 	&- \\\\

11     &1      &24.4 	&7.0 	&9.6 	&- 	&- 	&- 	&-\\\hline
\end{tabular}
\end{footnotesize}
\end{table*}

In this section we describe a set of tests that are related to the following scenarios:
\begin{itemize}
\item Scenario I: We assume that we have no information on the trajectory of the spacecraft, but wish to know its current position. 
\item Scenario II: We wish to know the position and velocity of the spacecraft at a specified time making use of a model for its trajectory (and assume that the observations are simultaneous)
\item Scenario III: We consider a more realistic scenario in which we wish to know the position and velocity of the spacecraft, but do not have simultaneous observations of multiple pulsars.
\item Scenario IV: We improve the realism of our modelling by assuming that the clock on the spacecraft is not perfect, but will drift over time.
\item Scenario V: We demonstrate the effect of a glitch event that occurs in a specific pulsar after the launch of the spacecraft.
\item Scenario VI: We describe how our algorithms can be applied for non-autonomous navigation.
\end{itemize}

These tests are summarised in Table~\ref{Tab:basic_summary} which indicates the phenomena that have been included.  In column order, the table gives 1) the simulation number, 2) the scenario being considered, 3) which of the two algorithms have been applied, 4) whether the timing models are extrapolated using the algorithms of \citet{deng2012optimal}, 5) whether the on-board observations are assumed to occur simultaneously or not, 6) if the on-board clock is assumed to drift, 7) whether drifts in the clock are accounted for in the analysis, 8) whether one pulsar has undergone a glitch event, 9) and 10) which of two procedures (described below) is used to account for the simulated glitch event and 11) whether the navigation is autonomous or not.

For each of our tests we create 100 realisations of the radio and X-ray data sets. As described below for each scenario we assume an incorrect guess for the spacecraft position (and velocity) for each realisation. We then use either algorithm 1 or algorithm 2 to determine the position (and, for algorithm 2, the velocity) of the spacecraft throughout the orbit for each realisation and record the discrepancy between the known simulated position (and velocity) and the measured values.  We wish to determine 1) whether, on average, we recover the simulated position (and velocity) without any bias and 2) what is the typical uncertainty in the measurement of the spacecraft position (and velocity) and 3) whether that uncertainty agrees with the parameter uncertainties obtained from the fitting procedure.  We choose 100 realisations in order to provide a sufficient number of samples to study the distribution, bias, scatter and error bar sizes of the resulting position estimates. 

We summarise our results in Tables~\ref{Tab:errorbar_result} and \ref{Tab:discrepancy_result}. For Table~\ref{Tab:errorbar_result}, the first column gives the identification number for the test. The second column gives the identification number of algorithms. The next three columns give the mean error bar in the position fitting of the 100 realisations for the last set of observations (when the spacecraft reaches Mars). The next three columns give the same for the spacecraft's velocity. The final column gives, for test 7, the mean error bar on the measurement of the spacecraft clock for the last set of observations. 

For Table~\ref{Tab:discrepancy_result}, the first column gives the identification number for the test. The second column gives the identification number of algorithms. The next three columns give the rms of the 100 measurements of the discrepancy between the simulated position and the measured position for the last set of observations. The next three columns give the same for the spacecraft velocity.  The final column gives, where necessary, the rms variation in the discrepancy between the spacecraft clock and the corrected spacecraft clock value. 

\subsection{Scenario I}
\label{Sect:Scenario1}
In this scenario we assume that we require the current position of the spacecraft without making any assumption about its trajectory.  We assume simultaneous pulsar observations and that the position of the spacecraft is known to within $\sim$\,100\,km.  We therefore take the known, simulated position of the spacecraft, but randomly change the position by a Gaussian random deviate scaled by 100\,km.   We then use algorithm 1 to fit for the spacecraft's position. 

\begin{figure*}[h]
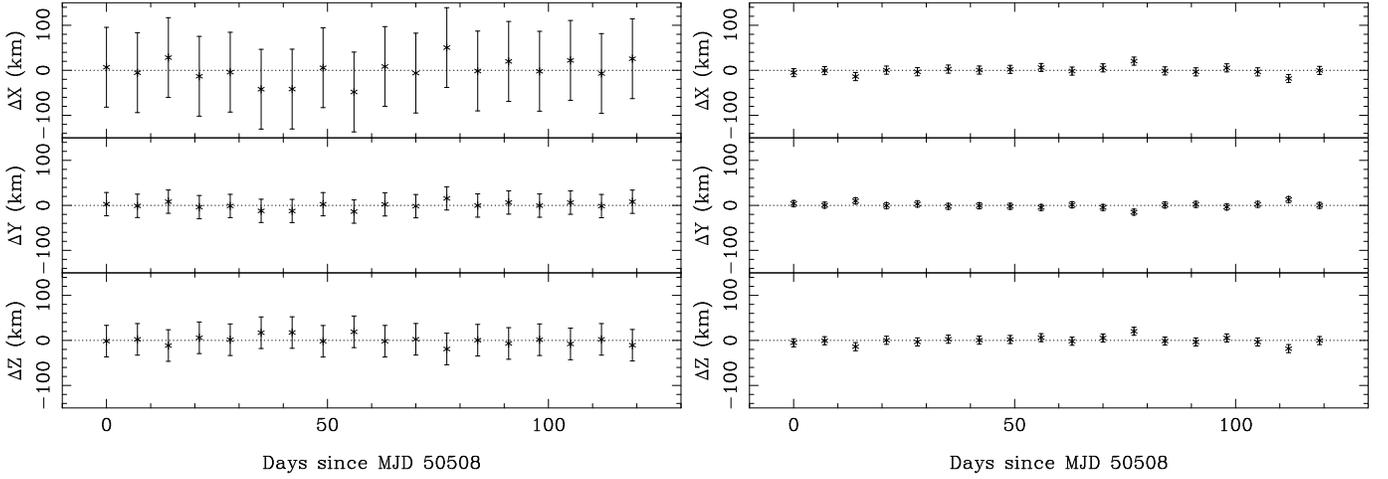

\centering
\includegraphics[height=9cm,angle=-90]{alg1_noextrap.ps}
\includegraphics[height=9cm,angle=-90]{alg1_extrap.ps}
\caption{The position fitting discrepancy using algorithm 1. The panels on the left hand side show the discrepancy in the x, y and z position determination without accounting for red timing noise (Test 1). The figure on the right hand side shows the discrepancy in the position determination after accounting for red timing noise (Test 2). \label{Fg:millisecondPosition}} 
\end{figure*}

For our first test, we do not update the timing models obtained from the ground-station and assume that the red timing noise can be ignored.  The discrepancy in the position determination as a function of time for one realisation of the data sets is shown in the left-hand panel of Figure~\ref{Fg:millisecondPosition}. The rms values of the discrepancy between the measured and simulated positions are 21.7, 6.4 and 8.5\,km in the x-, y- and z-axes respectively. The mean error bar size is 94.9, 27.7 and 37.6\,km respectively, which is significantly larger than the rms values. This occurs because the error sizes are scaled by the reduced-$\chi^2$ value of the fit.  If the red timing noise is not accounted for then the reduced-$\chi^2$ value is significantly greater than unity and the resulting error bars are large.

For our second test, we use the \citet{deng2012optimal} method to extrapolate the radio timing residuals to the time of the current observation. The discrepancy in the position determination as a function of time for one realisation of the data sets is shown in the right hand panel of Figure \ref{Fg:millisecondPosition}. The rms values of the discrepancy in the position are 24.6, 7.2 and 9.8\,km in the x-, y- and z-axes respectively.  In this test, the mean error bar size is consistent with the rms of the discrepancy of the position determination.

For our array of pulsars, the error in the position determination is significantly larger on the  x-axis than on the other axes. Ideally we would have chosen three pulsars that have identical ToA precision and lead to unit vectors, $\bf\hat{p}$, that are orthogonal, but this is not practical.

\begin{figure*}[h]
\centering
\includegraphics[height=9cm,angle=-90]{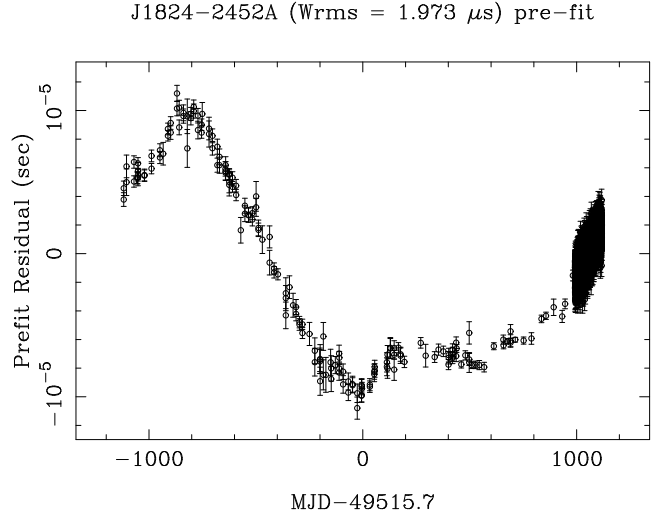}
\caption{One realisation of the simulated radio and X-ray observations for PSR~J1824$-$2452A. The simulated observations with small error bars are the radio observations. Those with large uncertainties, after MJD 50508, are the X-ray observations. \label{Fg:all-data}} 
\end{figure*}

It is clear that the extrapolation process is important.  However, it is never possible to predict the timing residuals perfectly over long time durations.  The error in the position determination will therefore increase as we extrapolate further into the future. We could not see any significant improvement in the rms of the discrepancy using the extrapolation procedure because 1) the time interval for the extrapolation process is very short for our simulated trajectory (only 4 months) and 2) the X-ray ToA uncertainties are much larger than the residuals induced by the red timing noise (we show the simulated radio and X-ray observations of PSR~J1939+2134 in Figure~\ref{Fg:all-data}).

\subsection{Scenario II}
\label{Sect:Scenario2}
In this scenario we assume that we wish to use the dynamics model to determine the position and velocity of the spacecraft.  We assume that pulsar observations are made every seven days (the time interval is the same with the navigational optical observing time interval of the Deep Space 1 spacecraft mission; \citealt{rayman2000results}) and that it takes 1\,d to carry out 24 observations in which all four pulsars are observed simultaneously. 

We start with initial guesses of the spacecraft position that are offset by 100\,km and velocity offsets of 0.01\,kms$^{-1}$ from the values given in Table~\ref{Tab:orbit-state}.    For the Monte-Carlo simulations we subsequently jitter these offsets by a Gaussian random deviate scaled by 10\,km in distance and 0.001\,kms$^{-1}$ in velocity. We use the simulated observations to improve our determination of the position of the spacecraft at the start of the X-ray observations using algorithm 2.  We then wait until the next X-ray pulsar observations, use the updated initial parameters to predict the current position and velocity of the spacecraft, form pulsar timing residuals and then re-fit for the position and velocity of the spacecraft at the start of the current observations. We repeat this procedure, obtaining a measurement of the spacecraft's position and velocity every seven days throughout the orbit. We only use the ground-based radio data taken before the launch of the spacecraft and the most recent 24 X-ray observations for each pulsar.

\begin{figure*}[h]
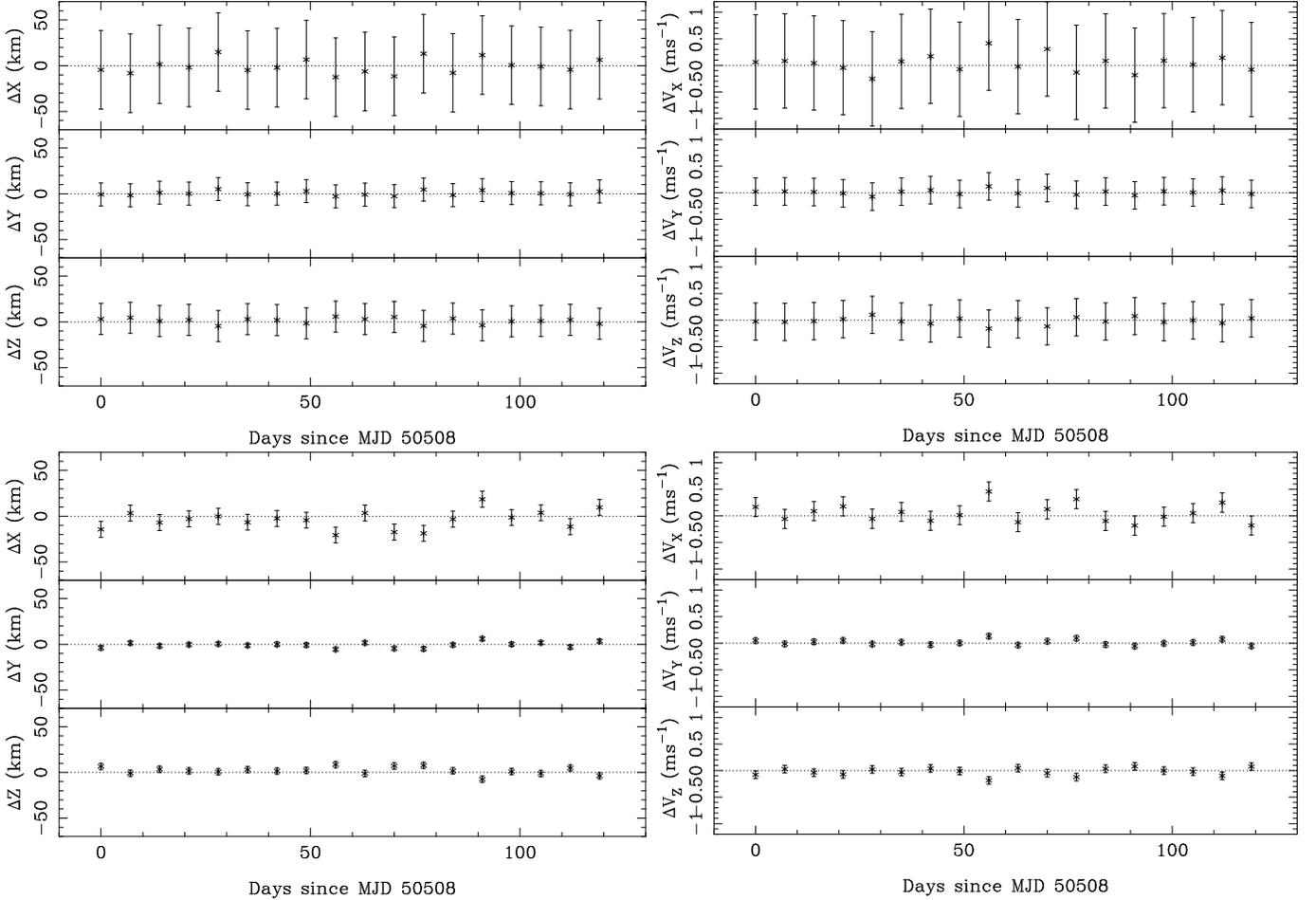

\centering
\includegraphics[height=9cm,angle=-90]{alg2_xyz_noextrap.ps}
\includegraphics[height=9cm,angle=-90]{alg2_vel_noextrap.ps}
\includegraphics[height=9cm,angle=-90]{alg2_xyz_extrap.ps}
\includegraphics[height=9cm,angle=-90]{alg2_vel_extrap.ps}
\caption{The position and velocity fitting discrepancy obtained using algorithm 2. The figures on the top show the discrepancy in the position and velocity determination without accounting for the red timing noise (Test 3). The figures on the bottom show the discrepancy in the position and velocity determination after accounting for the red timing noise (Test 4). The top, middle, bottom panels of the figures on the left hand side represent errors in the distance determinations, while those on the right hand side indicate the errors in the velocity determination. \label{Fg:result_scenario2}} 
\end{figure*}

In Figure~\ref{Fg:result_scenario2} we show the difference between the measurement of the spacecraft position and velocity compared with the simulated trajectory when we do not account for the pulsar red timing noise (upper panels; test 3) and after accounting for the noise (lower panels; test 4).  We again note that it is essential to extrapolate the timing residuals in order to account for the red timing noise when determining the error size. When we account for the red timing noise we recover the position with an rms discrepancy of 9.4, 2.8, 3.8\,km in the x-, y- and z-axes respectively and velocity with rms values for the discrepancy of 0.19\,ms$^{-1}$, 0.054\,ms$^{-1}$ and 0.073\,ms$^{-1}$ respectively. The mean error bar size is slightly smaller than the rms of the discrepancy because the dynamics model for algorithm 2 does not account for the solar radiation pressure. As expected, having an algorithm that can make use of multiple observations significantly improves the position determination compared with the results from algorithm 1.

\subsection{Scenario III: Non-simultaneous observations}

As defined, algorithm 1 can only be applied to simultaneous observations of multiple pulsars. If the observations are not simultaneous, then the algorithm needs to be updated to account for the motion of the spacecraft between observations.  For constant velocity, this is straightforward. However, as algorithm 2 uses a dynamical model to predict the spacecraft position and velocity, it can deal with non-simultaneous observations without modification.  The only assumption being that the dynamics model is accurate over the duration of the observations.   In order to test this situation we, as before, assume that  X-ray observations are made every seven days, but now we assume that 1\,d is allocated to obtain six non-simultaneous observations of each pulsar.   

For this scenario (test 5), we again start with initial guesses of the spacecraft position and velocity that are the same as in tests 3 and 4. We recover the position with rms values for the discrepancies of 16.2, 4.7, 6.5\,km in the x-, y- and z-axis respectively and velocity with rms values of the discrepancies of 0.36\,ms$^{-1}$, 0.10\,ms$^{-1}$ and 0.14\,ms$^{-1}$ respectively. The uncertainty here is larger than that obtained in scenario II as fewer observations are used for the same observing duration. However, this test demonstrates that simultaneous observations are not necessary for pulsar-based navigation.



\subsection{Scenario IV: The spacecraft clock}

Measurements of ToAs on the spacecraft will be made using an on-board clock. These ToAs will subsequently be converted to TCB.  Here we do not discuss the time transfer to TCB, but note that the onboard clock will drift over time and therefore errors in the clock will lead to incorrect determination of the barycentric arrival times. It is straightforward to update both of our algorithms to fit for an error in the spacecraft clock. Details are given in \ref{appendix:clk_fitting}.

\begin{figure*}[h]
\centering
\includegraphics[height=9cm, angle=-90]{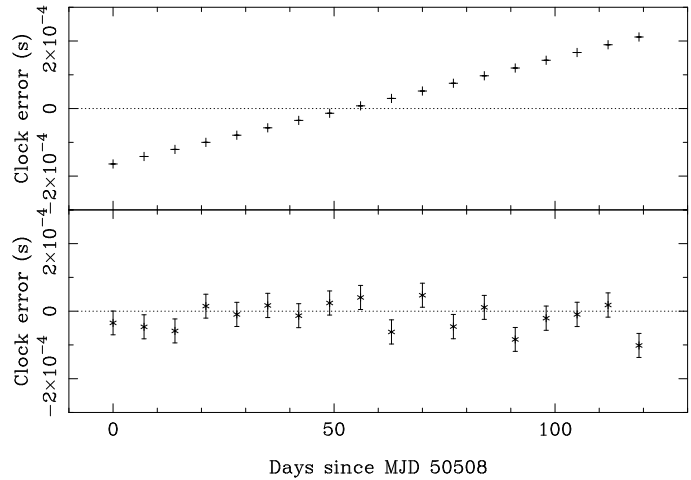}
\caption{The top panel shows for one realisation of the simulated error in the spacecraft clock. The bottom panel shows the difference between the actual, simulated clock error and the measured value using algorithm 2.\label{Fg:clock_fitting2}} 
\end{figure*}

The clock noise is likely to be a red-noise process \citep[and references therein]{hobbs2012development}.  However, it is also common for the clock to be corrected when the drift exceeds some threshold.  This correction will lead to discrete changes in the clock error.  Clock stability is continuing to improve and it is difficult to predict the expected behaviour of a clock on board a future spacecraft.  For simplicity we simply simulate a large red noise process for the error in the clock (one realisation of the simulated red noise process is shown in the top panel of Figure~\ref{Fg:clock_fitting2}).  We therefore demonstrate that our algorithms can account for clock noise that is much larger than that expected for a future mission.  Details for simulating the red noise are given in \ref{appendix:data_simulation}. If we do not take the clock noise into account then we obtain the position and velocity results given in Tables~\ref{Tab:errorbar_result} and \ref{Tab:discrepancy_result} as Test 6. We note that, in this case, the rms of the discrepancy is significantly larger than the mean error bar. 

For Test 7, we use the updated algorithm 2 to measure the clock error as part of the position (and velocity) determination procedure.  The discrepancy in the recovered clock error for one realisation is shown in the bottom panel of Figure~\ref{Fg:clock_fitting2}. All position and velocity measurements are consistent with the simulated values.

\subsection{Scenario V: Updating the pulsar timing ephemerides}

\label{sect:updatingephemerides}
A pulsar timing ephemeris can model the rotational properties of a pulsar with very high precision and accuracy.  However, as shown in Figure~\ref{Fg:dset1}, even millisecond pulsars exhibit unexplained red timing noise \citep{shannon2010assessing}. \citet{lyne2010switched} have recently shown that the red timing noise for many pulsars can be modelled by assuming that the pulsar's spin-down rate flips at pseudo-random intervals between two states.  It is not clear whether this phenomenon occurs also for millisecond pulsars, but \citet{manchester2013} have shown that more than half of the PPTA pulsars exhibit low-frequency red timing noise.  Pulsars are also known to exhibit ``glitch events'' in which the pulse period suddenly decreases.  Many glitches are now known for young pulsars (e.g., \citealt{espinoza2011study}), which seriously limits the use of young pulsars for navigational purposes. One millisecond pulsar in our sample, PSR~J1824$-$2452A, has also been known to glitch \citep{cognard2004micro}. The event was characterised by a rotational frequency step of 3\,nHz.  

In the scenarios above, we have assumed that ground-based observations are obtained until the launch of the spacecraft.  After launch, the spacecraft runs autonomously and no contact between the spacecraft and the ground is made.  We also assumed that the pulsar timing ephemerides do not get updated after the launch. 

For this section, we make the assumption that a glitch event occurred in PSR~J1824$-$2452A just after the launch of the spacecraft. The parameters for the glitch are given in Table \ref{tab:glitch_infor}. It may be possible to obtain sufficient ground-based radio data to model the glitch event and include it in the pulsar timing model.  If this is possible then the effect of the glitch event can be completely mitigated. However, it is likely that the recovery from the glitch event is indistinguishable from the red timing noise and therefore the event cannot be fully modelled in the timing ephemeris.  

We describe two methods that can be used to deal with the glitch event.  In the first method we  assume that the resulting timing residuals can be modelled as red timing noise.  We note that the glitch introduces non-stationary noise into residuals.  For small glitches, \citet{deng2012optimal} demonstrated that this non-stationarity does not significantly affect the extrapolation procedure. In the second model we assume that the spacecraft has no contact with the ground since launch and therefore only the X-ray data can be used to model the timing residuals induced by the glitch.

\begin{table*}[!t]
\caption{The parameters for the simulated glitch in PSR~J1824$-$2452A \label{tab:glitch_infor}}
\centering
\begin{footnotesize}
\begin{tabular}{ll} \hline
Parameter 				&Value \\\hline
Glitch Epoch (MJD)			&50508\\
Glitch Frequency (Hz)			&3.12$\times10^{-9}$\\
Glitch Frequency Derivative (Hz/s)	&5.9$\times10^{-18}$\\\hline
\end{tabular}
\end{footnotesize}
\end{table*}

\begin{figure*}[h]
\centering
\includegraphics[height=9cm,angle=-90]{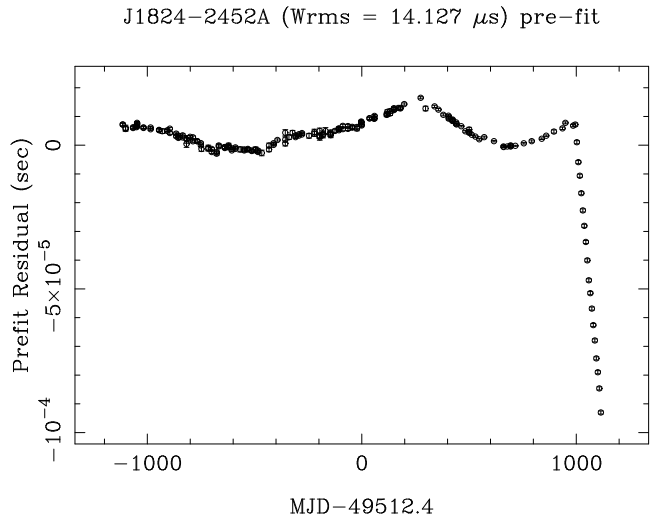}
\caption{One realisation of the simulated data set that includes a glitch event for PSR J1824$-$2452A at the time of launch.  The glitch event is observed as the sudden change in slope of the timing residuals at MJD 50508. \label{Fg:glitch_data}} 
\end{figure*}

The simulated ground data for PSRs J0437$-$4715, J1939$+$2134 and J2124$-$3358 are identical to the simulations in Section~\ref{sect:ground-data}.  For PSR J1824$-$2452A we have simulated the ground data until the end of the mission to Mars and have added a  glitch event that occurred at the time of the launch.  In Figure \ref{Fg:glitch_data}, we show the residuals for PSR~J1824$-$2452A relative to a model that does not include a glitch.

\begin{figure*}[h]
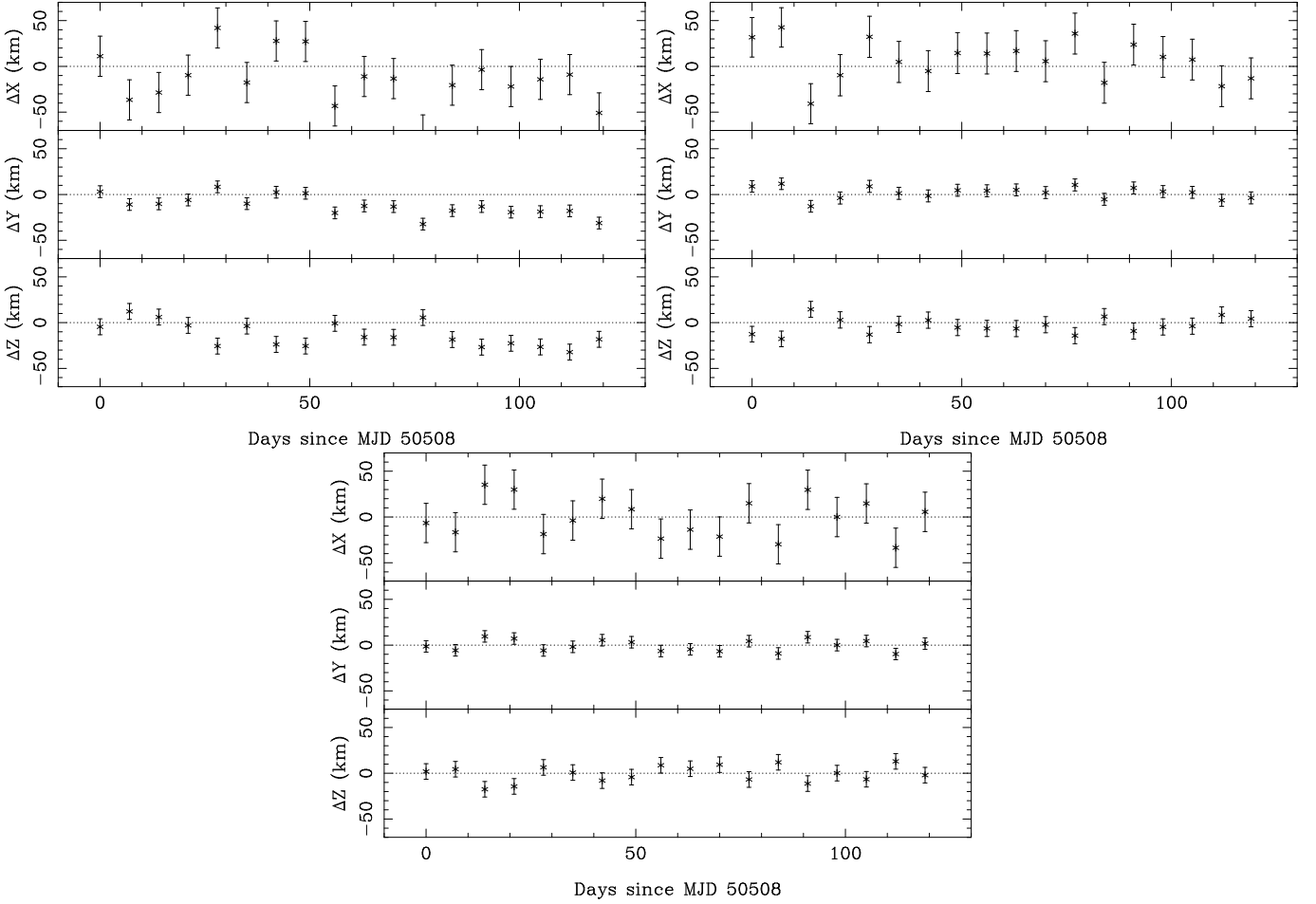

\centering
\includegraphics[height=9cm,angle=-90]{glitch_comparison.ps}
\includegraphics[height=9cm,angle=-90]{glitch_radio.ps}\\
\includegraphics[height=9cm,angle=-90]{glitch_xray.ps}
\caption{Position determination using algorithm 1 when PSR J1824$-$2452A has undergone a glitch event. The figure on the top-left shows the result without accounting for the glitch event (Test 8). The figure on the top-right shows the result with the ephemerides updated to account for the glitch using radio observations (Test 9). The lowest figure shows the result after the ephemerides have been updated using X-ray observations (Test 10). For each figure, the top, middle and the bottom panel represents the discrepancy in the position determination for the x-, y- and z-axes respectively.\label{Fg:glitch_all}} 
\end{figure*}

For comparison, we first (Test 8) apply algorithm 1 to this new data set without accounting for the glitch event.  As shown in the left hand side of Figure~\ref{Fg:glitch_all}, the position discrepancy increases with time and the rms of the discrepancy in the position determination for 100 realisations is 31.2, 22.2, 31.3\,km in the x-, y- and z-axes respectively. 

\subsubsection{Method 1: Ephemerides updated using radio observations}
\label{sect:ephemerides2}
For this test (Test 9), we assume that radio observations are available up to seven days before each X-ray observation.  Before processing the X-ray data, we refit the radio observations to obtain a new timing model for PSR J1824$-$2452A, determine a new spectral model and extrapolate to the X-ray observations. With this method (middle panel of Figure~\ref{Fg:glitch_all}), the position discrepancy does not increase with time and the rms of the discrepancy in the position determination for 100 realisations is 23.3, 6.8, 9.3\,km in the x-, y- and z-axis respectively. The results are promising and this technique is applicable for real-time navigation, but requires that an updated radio ephemeris can be transmitted to the spacecraft.



\subsubsection{Method 2: Ephemerides updated using X-ray observations}

For this test (Test 10), we again assume that the spacecraft must be navigated autonomously and we update the timing model using the X-ray observations.  In order to do this we use all four pulsars to determine the spacecraft position at a given time. We then include the X-ray observations when we form the timing residuals (the ToA uncertainties on the X-ray arrival times are increased to account for the error in the position determination). For the pulsar that has glitched, we then re-fit for the pulsar timing model using the radio data and all the previously determined X-ray ToAs.  This new timing model is subsequently used when processing the next set of X-ray observations.  If the error induced by the distance determination is comparable to or smaller than the original ToA uncertainty then,  as shown in Test 10, we can successfully recover the spacecraft position.    As this method does not require any radio observations after launch, it  provides a completely autonomous navigation method for the spacecraft.  However, it would be necessary for the onboard system to identify that a glitch has occurred and the precision with which the position of the spacecraft can be determined must be similar (or better) than the ToA uncertainty on the X-ray observations.

\subsection{Scenario VI: Non-autonomous navigation}

For most of this paper we have assumed that the spacecraft must be navigated autonomously.   This is a stringent condition that is not required for current interplanetary spacecraft.  If the spacecraft computers can be updated from Earth then the most up-to-date pulsar ephemerides from the ground station can be uploaded and used in the determination of the spacecraft's position.  This therefore removes most issues resulting from glitch events or pulsar red timing noise.  To demonstrate this we carry out a final set of simulations which are identical to Test 2 (i.e., using algorithm 1), but the ground station data is assumed to be available up to 1 hour before the required position determination.  A new timing model that includes an interpolation of the red timing noise for each pulsar is obtained from these ground station observations and uploaded to the spacecraft. The rms of the discrepancy between the measured and simulated positions for this non-autonomous navigation are 24.4, 7.0 and 9.6\,km in the x-, y- and z-axes respectively.  We note that there is no significant change with Test 2 because the timescale of extrapolation is relatively short, however, this non-autonomous procedure significantly reduces the necessary computational resources on the spacecraft.


\subsection{The use of \textsc{tempo2} for real-time applications}

Our algorithms rely on the \textsc{tempo2} software package to form barycentric times-of-arrival and to carry out the necessary fits to determine the spacecraft position.  This software package has not been designed for spacecraft navigation and the current software is not suited to spacecraft missions.  It currently is not memory efficient and requires significant processing power.  \textsc{Tempo2} is designed for astronomical applications and therefore includes numerous features that are not necessary for navigation purposes. The software also makes use of the following external libraries that are not part of the standard \textsc{tempo2} distributions:

\begin{itemize}
\item The ``Standards of Fundamental Astronomy" software library that is provided by the International Astronomical Union
\item A file providing parameters for the Earth's orientation from the International Earth Rotating and Reference Systems service
\item Various files that correct the time measured at an observatory to Terrestrial Time.  The most commonly used files are available from the Bureau International des Poids et Mesures.
\item A Solar System planetary ephemeris.  Most users of \textsc{tempo2} use the Jet Propulsion Laboratory planetary ephemerides.  However, \textsc{tempo2} can also use the European INPOP planetary ephemerides.
\end{itemize}

The purpose of this paper is to show that pulsar-based navigation is practical and that existing software implements all the necessary algorithms with sufficient precision.  However, any actual mission would require a complete re-write of the software.   Non-autonomous navigation is simpler as \textsc{tempo2} (or equivalent software) can run on standard work-stations at the ground-station.  However, even for this application, we would recommend a simplification of the \textsc{tempo2} software.

\section{Conclusion}

It has long been known that observations of pulsars can be used to measure the position and velocity of a spacecraft. We have presented new results based on observations of millisecond pulsars that have been observed as part of the Parkes Pulsar Timing Array project.  The algorithms that we have described rely on up-to-date knowledge of the rotational parameters for each pulsar at the time of launch via a ground-based radio telescope monitoring programme. The precision and accuracy with which the spacecraft's position and velocity can be determined depends on the number of pulsars observed, the number of pulsar observations, the precision with which pulse arrival times can be measured and the amount of time that the timing model needs to be extrapolated.  With our ensemble of realistic measurements of four millisecond pulsars we have shown that the position of a spacecraft on a trajectory from Earth to Mars can be determined to $\sim$\,20\,km. Note that the exact precision achievable depends upon exactly how the pulsars are observed and whether the on-board clock can easily be corrected for long-term drift, or not.  We also note that, with the four pulsars in our sample, it is possible to determine the spacecraft position in two  barycentric coordinates (y and z) more precisely than in the third coordinate.

For now we have not attempted to deal with pulse ambiguities that occur when the error in the estimated spacecraft position is significantly incorrect (i.e., when the initial position estimate is less precise than $cP_{\rm fast}/2$). For our pulsar sample the fastest spinning pulsar has a period of 1.5\,ms and so our algorithms rely on having the initial guess of the position known to better than $\sim$225\,km.  We are currently considering methods to solve this problem, but there are a few straight-forward possibilities:

\begin{itemize}
\item Include more pulsars in the sample that have slower spin rates.  The most slowly spinning pulsars can provide the initial estimate of the spacecraft position which then gets improved as the data from more pulsars are added.
\item Use another navigational method to obtain the initial estimate of the spacecraft position. Combining an INS with an on-board camera should provide a sufficient estimate of the spacecraft position to enable the use of the algorithms described in this paper.  We note that the final position estimate is likely to be dominated by the precision of the pulsar measurements, but the INS and camera are essential in obtaining the initial position estimate.
\end{itemize}


Pulsar navigation provides a completely autonomous means for navigating spacecraft. With future developments in telescope design, clock stability and on-board computing power, it is likely that spacecraft could be navigated with sub-km precision.






\section*{Acknowledgments} 

The Parkes radio telescope is part of the Australia Telescope which is funded by the Commonwealth of Australia for operation as a National Facility managed by CSIRO. 

XPD acknowledges support from the China Scholarship Council (CSC) which provided funding to study in Australia for one year. CSC is a non-profit institution affiliated with Ministry of Education of the P. R. China. It is entrusted by the Chinese Government with the responsibilities of managing the State Scholarship Fund and other related affairs. It sponsors Chinese citizens to pursue study abroad and international students to study in China. GH is the recipient of an Australian Research Council QEII Fellowship (project \#DP0878388) and acknowledges support from the National Natural Science Foundation of China (NSFC) \#10803006 and \#11010250. XPY acknowledges support from the National Natural Science Foundation of China (NSFC) \#U1231120 and the Fundamental Research Funds for the Central Universities (XDJK2012C043).

\section*{References} 
\bibliographystyle{model2-names}
\bibliography{myrefs}





\appendix

\section{\textsc{tempo2} plugins usage instructions}
\subsection{Plugins for data simulation}
\label{appendix:data_simulation}
\textsc{tempo2} uses the flag \verb|STL_FBAT| to identify the X-ray observations taken by the X-ray telescope on the spacecraft and \verb|-telx|, \verb|-tely| and \verb|-telz| to identify the position of spacecraft in SSB coordinates (measured in lightseconds). The flag \verb|-mjdTT| is used to record the TT time correspond with the current position. One example of the X-ray observation entry is as follow:
\begin{verbatim}
stl_J0437-4715 1440.00000000 50531.75072026088205490 58.00000 STL_FBAT 
-telx -639.287517 -tely 67.604940 -telz 30.659395 -mjdTT 50531.75071972
\end{verbatim}

\begin{itemize}
\item{\textsc{formIdeal}}

The \textsc{formIdeal} plugin provides us with an ``ideal'' arrival time for each observation that is perfectly modelled by the pulsar timing model.  They can be formed as follows
\begin{verbatim}
> tempo2 -gr formIdeal -f mypar.par mytim.tim 
\end{verbatim}
where \verb|mypar.par| and \verb|mytim.tim| are the parameter and arrival time files for our simulation respectively. This plugin will produce an ideal arrival time file with the name \verb|mytim.tim.sim|.

\item{\textsc{addGaussian}}

\textsc{addGaussian} will produce a file \verb|mytim.tim.sim.addGauss| that contains a list of the time offsets relating to the measurement uncertainty.  This plugin can be used as follows:
\begin{verbatim}
> tempo2 -gr addGaussian -f mypar.par mytim.tim.sim 
\end{verbatim}
where \verb|mytim.tim.sim| is output by the \textsc{formIdeal} plugin.

\item{\textsc{addRedNoise}}

\textsc{addRedNoise} will produce a file \verb|mytim.tim.sim.addRedNoise| that contains the time offsets relating to the red timing noise. This plugin can be used as follows:
\begin{verbatim}
> tempo2 -gr addRedNoise -f mypar.par mytim.tim.sim 
  -fc fc -a a -Pyr3 Pyr3
\end{verbatim}
Where \verb|fc| is the same with the $f_c$ described in Equation \ref{eq:spectralModel}, \verb|a| is $-\alpha$ in Equation \ref{eq:spectralModel} and \verb|Pyr3| is $0.5{P_0}$ in Equation \ref{eq:spectralModel}.

\item{\textsc{createRealisation}}

\textsc{createRealisation} will add the various time delays to the ideal arrival time file  \verb|mytim.tim.sim| to form the final arrival time file \verb|mytim.tim.sim.real|:
\begin{verbatim} 
> tempo2 -gr createRealisation -f mytim.tim.sim 
  -corr mytim.tim.sim.addGauss -corr mytim.tim.sim.addRedNoise
\end{verbatim}

\item{\textsc{addGWB}}
We use the {\textsc{addGWB}} plugin to simulate the red noise that represent variations in the on-board clock:
\begin{verbatim}
> tempo2 -gr addGWB -f mypar.par mytim.tim.sim.real 
  -dist 1 -alpha -0.6666 -gwamp 5e-13-ngw 1000
\end{verbatim}

\end{itemize}

\subsection{Plugins for navigation}
\label{appendix:navigation}
\begin{itemize}
\item{Plugin for algorithm 1}

We use the global fitting procedures within \textsc{tempo2} to fit for the error in the telescope position:
 
\begin{verbatim}
> tempo2 -f mypar1.par mytim1.tim -f mypar2.par mytim2.tim 
  -f mypar3.par mytim3.tim -f mypar4.par mytim4.tim
  -global global.par -fitfunc global
\end{verbatim}

where \verb|global.par| has the following form:
\begin{verbatim}
STEL_DX 2 2
TEL_DX1 mjd1 0 0
TEL_DX2 mjd2 0 0

STEL_DY 2 2
TEL_DY1 mjd1 0 0
TEL_DY2 mjd2 0 0

STEL_DZ 2 2
TEL_DZ1 mjd1 0 0
TEL_DZ2 mjd2 0 0

STEL_CLK_OFFS 2 2
TEL_CLK_OFFS1 mjd1 0 0
TEL_CLK_OFFS2 mjd2 0 0
\end{verbatim}

The \verb/mjd1/ and \verb/mjd2/ give the time interval for the fitting, the \verb/STEL_DX/, \verb/STEL_DY/, \verb/STEL_DZ/ and \verb/STEL_CLK_OFFS/ tell \textsc{tempo2} to fit for the error in the position of the spacecraft and the variation in the clock.
 
\item{Plugin for algorithm 2}

For algorithm 2, we use the \textsc{navOrbit} plugin:
\begin{verbatim}
> tempo2 -gr navOrbit -f mypar1.par mytim1.tim 
  -f mypar2.par mytim2.tim -f mypar3.par mytim3.tim 
  -f mypar4.par mytim4.tim -global global.par 
  -fitfunc global
\end{verbatim}
where \verb|global.par| contains the basic information necessary for the simulation. The template for \verb|global.par| is as follows: 
\begin{verbatim}
TELEPOCH mjd1

TEL_X0 X0 2
TEL_Y0 Y0 2
TEL_Z0 Z0 2

TEL_VX VX 2
TEL_VY VY 2
TEL_VZ VZ 2

STEL_CLK_OFFS 2 2
TEL_CLK_OFFS1 mjd1 0 0
TEL_CLK_OFFS2 mjd2 0 0
\end{verbatim}

Where \verb/mjd1/ and \verb/mjd2/ are the start and end times of the observations.

\end{itemize}
\subsection{Plugins for data analysis}
\label{appendix:data_analysis}
\begin{itemize}
\item{\textsc{spectralModel}}

{\textsc{spectralModel}} plugin can be used to obtain a spectral model for a given pulsar data set:
\begin{verbatim}
> tempo2 -gr spectralModel -f mypar.par mytim.tim 
\end{verbatim}

\item{\textsc{autoSpectralFit}}

This plugin can model the spectral of red timing noise automatically. The result is written into the file \verb/mymodel.model/.
\begin{verbatim}
> tempo2 -gr autoSpectralFit -f mypar.par mytim.tim 
\end{verbatim}

\item{\textsc{interpolate}}

This plugin can be used to interpolate or extrapolate the pulsar timing residuals.

\begin{verbatim}
> tempo2 -gr interpolate -f mypar.par mytim.tim 
    -a A -fc fc -alpha alpha
\end{verbatim}
\end{itemize}

\section{Algorithms updating for clock}
\label{appendix:clk_fitting}
In order to take the clock error into account, we have to update Equation \ref{eq:basic} to

\begin{equation}
\label{eq:basic_clk}
R =\frac{1}{c}{\bf\hat{p}}^\mathrm{T}\delta{\bf{r}}_\mathrm{SSB}+\delta{t}_\mathrm{clk}
\end{equation}
where $\delta{t}_\mathrm{clk}$ is the error in the clock at a given time.

For algorithm 1, the Equation \ref{eq:basic_clk} can be modified to
\begin{equation}\label{eq:algorithm1_clock}
\begin{split}
R_i &=\frac{1}{c}{\bf\hat{p_i}}^\mathrm{T}\delta{\bf{r}}_\mathrm{SSB}+\delta{t}_\mathrm{clk}\\
&=\frac{1}{c}\left[{\bf\hat{p_i}}^\mathrm{T}~~1\right]\left[\delta{\bf{r}}_\mathrm{SSB}^\mathrm{T}~~c\delta{t}_\mathrm{clk}\right]^\mathrm{T}.
\end{split}
\end{equation}

For algorithm 2, Equation \ref{eq:Rel-final} can be updated to
\begin{equation}
\begin{split}
{R_\mathrm{ij}}&=\frac{1}{c}[{\bf\hat{p}}_\mathrm{i}^T~~{\bf\hat{p}}_\mathrm{i}^T{t_\mathrm{j_0}}]\delta{{\bf{x}}_\mathrm{SSB_0}}+\delta{t}_\mathrm{clk_0}\\
&=\frac{1}{c}[{\bf\hat{p}}_\mathrm{i}^T~~{\bf\hat{p}}_\mathrm{i}^T{t_\mathrm{j_0}}~~1][\delta{{\bf{x}}_\mathrm{SSB_0}}^\mathrm{T}~~c\delta{t}_\mathrm{clk_0}]^\mathrm{T}\\
\end{split}\label{eq:Rel-final_clock}
\end{equation}
where $\delta{t}_\mathrm{clk_0}$ is the error in the clock at $t_0$. 
\end{document}